\documentclass[journal,twoside,web]{IEEEtran}
\usepackage{cite}
\usepackage{amsmath,amssymb,amsfonts}
\usepackage{algorithmic}
\usepackage{graphicx}
\usepackage{textcomp}
\usepackage{wrapfig,colortbl}

\usepackage{diagbox}
\usepackage{makecell}
\usepackage[T1]{fontenc}
\usepackage{soul}
\usepackage[utf8]{inputenc}
\usepackage{graphicx}
\usepackage{multicol}
\usepackage{multirow}
\usepackage{amsmath,amsfonts,amssymb}
\usepackage{mathtools}
\usepackage{gensymb}
\usepackage{graphicx}
\usepackage{float}
\usepackage{algorithm}
\usepackage{algorithmic}

\usepackage{enumerate}
\usepackage{url}
\usepackage{multirow}
\usepackage{lipsum}
\usepackage{color}
\usepackage[colorlinks=true, allcolors=blue]{hyperref}
\usepackage{hhline} %
\usepackage{stfloats}

\DeclareMathOperator*{\argmin}{arg\,min}
\ifCLASSINFOpdf
\else
\fi

\begin{document}

\title{Investigating the Use of Traveltime and Reflection Tomography for Deep Learning-Based Sound-Speed Estimation in Ultrasound Computed Tomography}
\author{Gangwon~Jeong,
        Fu~Li,
        Trevor~M.~Mitcham,
        Umberto~Villa,
        Nebojsa Duric,
        and~Mark~A.~Anastasio
\thanks{This work was supported in part by NIH under Award R01EB028652, in part by the National Science Foundation’s Supercomputer Centers Program, in part by the State of Illinois, in part by the University of Illinois, and in part by the National Science Foundation. Computational Resources for this work were granted to the authors by the Delta Research Computing Project under Award OCI 2005572.}

\thanks{G. Jeong, F. Li, and M. Anastasio are with the Department of Bioengineering, University of Illinois Urbana-Champaign, Urbana, IL, 61801 USA. e-mail: \url{gangwon2@illinois.edu}, \url{fuli2@illinois.edu}, and \url{maa@illinois.edu.}}
\thanks{T. Mitcham and N. Duric  are with Department of Imaging Sciences, University of Rochester Medical Center, NY, 14642, USA. e-mail: \url{Trevor_Mitcham@URMC.Rochester.edu} and \url{Nebojsa_Duric@URMC.Rochester.edu.}}
\thanks{U. Villa is with the Oden Institute for Computational Engineering and Sciences, The University of Texas at Austin, Austin, TX, 78712 USA. e-mail: \url{uvilla@oden.utexas.edu.}}}
\maketitle

\begin{abstract}
Ultrasound computed tomography (USCT) quantifies acoustic tissue properties such as the speed-of-sound (SOS). 
Although full-waveform inversion (FWI) is an effective method for accurate SOS reconstruction, it can be computationally challenging for large-scale problems. 
Deep learning-based image-to-image learned reconstruction (IILR) methods can offer computationally efficient alternatives.
This study investigates the impact of the chosen input modalities on IILR methods for high-resolution SOS reconstruction in USCT. 
The selected modalities are traveltime tomography (TT) and reflection tomography (RT), which produce a low-resolution SOS map and a reflectivity map, respectively.
These modalities have been chosen for their lower computational cost relative to FWI and their capacity to provide complementary information: TT offers a direct SOS measure, while RT reveals tissue boundary information.
Systematic analyses were facilitated by employing a virtual USCT imaging system with anatomically realistic numerical breast phantoms. 
Within this testbed, a supervised convolutional neural network (CNN) was trained to map dual-channel (TT and RT images) to a high-resolution SOS map.
Single-input CNNs were trained separately using inputs from each modality alone (TT or RT) for comparison.
The accuracy of the methods was systematically assessed  using normalized root mean squared error (NRMSE), structural similarity index measure (SSIM), and peak signal-to-noise ratio (PSNR).
For tumor detection performance, receiver operating characteristic analysis was employed.
The dual-channel IILR method was also tested on clinical human breast data. 
Ensemble average of the NRMSE, SSIM, and PSNR evaluated on this clinical dataset were 0.2355, 0.8845, and 28.33 dB, respectively.
\end{abstract}
\bigskip
\IEEEpeerreviewmaketitle

\section{Introduction}
\label{sec:introduction}
\IEEEPARstart{U}{ltrasound} computed tomography (USCT) has emerged as a promising non-invasive medical imaging modality, demonstrating great potential in various clinical applications \cite{duric2007detection, sandhu2015frequency, duric2014clinical}. 
USCT image reconstruction methods can produce accurate estimates of the acoustic properties of tissue, including speed-of-sound (SOS), acoustic attenuation (AA), and density.
These maps are useful parameters for tissue characterization and provide accurate information for breast cancer diagnosis, treatment planning, and prevention \cite{andre2013clinical, kratkiewicz2022ultrasound, littrup2021multicenter, duric2020novel, wiskin20173, wiskin2019quantitative}.
This study specifically focuses on the high-resolution reconstruction of the SOS distribution, which is strongly correlated with tissue density and has proven useful in enhancing the sensitivity and specificity of breast cancer detection \cite{duric2007detection, sak2011relationship}.


Full-waveform inversion (FWI) has emerged as an effective method for reconstructing accurate and high-resolution SOS maps \cite{wang2015waveform,matthews2017regularized,li2023forward,ali2022open, li20233d}. 
By modeling the multiple scattering, dispersion, and diffraction effects present in the measurements, FWI can provide improved SOS estimates compared to those reconstructed by ray-based methods \cite{anis2014investigation,li2009vivo,hormati2010robust,ali2019open}.
Nevertheless, the computational cost of FWI is much higher when compared to ray-based approaches due to the necessity of solving wave equations. 
In particular, the computational feasibility of FWI is often limited when solving large-scale problems such as 3D reconstructions \cite{lucka2021high}. 
In order to effectively manage computationally demanding tasks, it is necessary to utilize high-end computing infrastructure that possibly relies on graphics processing units (GPUs) \cite{gokhberg2016full, guasch2020full}. 
However, the hardware cost to attain practical speeds for high-resolution SOS reconstructions on these systems can be significant.
Therefore, there remains an important need to develop computationally efficient reconstruction methods for USCT that can facilitate its widespread translation, which may also enable reduced-cost USCT systems that don't require extensive computing hardware for use in low-resource settings.

Learned reconstruction methods hold potential to reduce image reconstruction times and potentially improve image quality, with a primary focus on measurement-to-image learned reconstruction (MILR) and image-to-image learned reconstruction (IILR) approaches. 
MILR methods \cite{wu2018inversionnet, prasad2022deepuct, lozenski2024learned} map measurement data directly to an estimate of the desired object property by use of a neural network, which is often defined as a convolutional neural network (CNN).
However, due to the high dimensionality of the measurement data, to-date, these methods have been mostly limited to relatively small-scale problems. 
Also, their performance for reconstructing objects of clinical relevance remains largely unexplored.
Alternatively, IILR methods employ a neural network to map a preliminary estimate of the object property, typically computed by use of a computationally efficient method, to a high quality one \cite{jin2017deep, hauptmann2020unreasonable}.
Intrinsically, IILR methods can utilize cross-correlations between the input and target images, thereby enhancing scalability and efficiency, and can potentially decrease the complexity of the required network.
Despite their potential, IILR methods for SOS reconstruction in USCT have been relatively unexplored.
Additionally, the use of multi-modality inputs to an IILR method for SOS estimation remains to be systematically explored.

This work investigates the use of traveltime tomography (TT) and reflection tomography (RT) to produce images that are mapped to a high quality estimate of the SOS distribution by an IILR method.
Specifically, a low-resolution estimate of the SOS distribution produced by TT and a reflectivity image generated by RT serve as inputs to the IILR method.
One motivation behind using these dual-modality inputs is that they provide complementary information: TT provides quantitative information on the SOS distribution of the objects, while the reflectivity image produced by RT yields high-resolution information about tissue boundaries.
The reported study seeks to investigate the extent to which an IILR method can stably and accurately map these two images to a single high-resolution estimate of the SOS distribution, with a computational burden that is greatly reduced as compared to FWI.

The use of the dual-modality inputs for IILR method is systematically assessed through virtual imaging studies and tested on clinical human breast data.
The virtual imaging studies employ a 2D virtual USCT imaging system with anatomically accurate numerical breast phantoms to analyze the impact of dual-modality inputs on high-resolution SOS estimation.
Within this test bed, both low-resolution SOS maps and reflectivity maps are reconstructed using TT and RT, serving as inputs to a CNN.
Using a supervised approach, the CNN is trained using the standard mean squared error (MSE) loss, where the target output is the ground truth SOS map of the numerical breast phantom.
These virtual imaging studies consist of three simulation-based studies.
The first study evaluates the dual-channel CNN both qualitatively and quantitatively.  
To understand the impact of each modality, separate single-channel CNN mappings are trained on either TT or RT images, and their results are compared with those of the dual-modality CNN.
The second study assesses the generalizability of the dual-channel CNN to data not present in the training dataset, particularly data that might be considered out-of-distribution.
The third study aims to investigate the impact of the rare occurrence of tumors within the training dataset on the dual-channel CNN. 
This involves assessments of the dual-channel CNN's reconstruction accuracy in tumor regions and its performance in tumor detection tasks.
Finally, to complement these virtual imaging studies and assess the method's performance under clinical conditions, a preliminary in-\emph{vivo} study using  experimental breast USCT  data is  conducted.

The remainder of this article is organized as follows.
Section \ref{sec:backgrounds} reviews the USCT imaging model, the bent-ray-based traveltime tomography (BRTT) to reconstruct low-resolution SOS maps, and the delay-and-sum (DAS)-based reflection tomography method for reconstructing reflectivity images.
Section \ref{sec:methods} introduces the dual-channel IILR method and the fine-tuning approach to improve tumor reconstruction accuracy.
Section \ref{sec:numerical studies} describes the design of three virtual imaging studies: 1) The qualitative and quantitative analysis of the dual-modality IILR is conducted; 2) The generalizability of the dual-channel IILR to reconstruct object types that are not encountered during training is assessed; 3) The accuracy of SOS estimates in tumor regions and tumor detection performance of the dual-channel IILR are evaluated.
Section \ref{sec:results} presents the findings of the three virtual imaging studies.
Section \ref{invivo} describes a preliminary study in which the dual-modality IILR method is demonstrated with clinical USCT data.
Section \ref{sec:discussion} provides the discussion derived from these findings.
Finally, Section \ref{sec:conclusions} provides a summary of the work.

\section{Background}
\label{sec:backgrounds}
This section presents an overview of the USCT imaging model and basic principles of BRTT and DAS-RT, which are used to generate inputs for the IILR method described in Section \ref{sec:methods}.
\subsection{USCT imaging model}
\label{sec:backgrounds:usmodel}
In USCT, an emitting transducer produces an acoustic wavefield that propagates   through the object being imaged.
The spatiotemporal source function that excites the emitting transducer is denoted as $s(\mathbf{r},t)\in \mathbb{L}^2(\mathbb{R}^3 \times [ 0,T)) $, where $\mathbf{r}\in\mathbb{R}^3$ represents the spatial coordinate and $t\in [0,T)$ is the time coordinate. 
The quantity $T$ represents the acquisition time.
The resulting pressure wavefield distribution is denoted by $p(\mathbf{r},t)\in \mathbb{L}^2(\mathbb{R}^3 \times [ 0,T))$.
The acoustic wave propagation in an unbounded and lossy media can be modeled by the following three-coupled first-order partial differential equations \cite{tabei2002k,tillett2009k, treeby2010modeling}:
\begin{align} \label{eq:waveeqs}
\frac{\partial }{\partial t} \mathbf{u}(\mathbf{r}, t) &= -\frac{1}{\rho_0(r)}\nabla p(\mathbf{r},t), \\
 \label{eq:waveeqs2}
\frac{\partial}{\partial t} \rho(\mathbf{r}, t) &= -\rho_0(\mathbf{r})\nabla \cdot \mathbf{u}(\mathbf{r},t) + 4\pi\int_{0}^{t}dt'\, s(\mathbf{r},t'), \\
 \label{eq:waveeqs3}
\begin{split}
p(\mathbf{r},t) &= c^2(\mathbf{r})\Biggl[1 + \tau(\mathbf{r})\frac{\partial }{\partial t}\left(-\nabla^2\right)^{y/2-1} \\
&\quad + \eta(\mathbf{r})\left(-\nabla^2\right)^{(y+1)/2-1} \Biggr]\rho(\mathbf{r},t),
\end{split}
\end{align}
where $\mathbf{u}(\mathbf{r},t)$, $c(\mathbf{r})$, $\rho(\mathbf{r},t)$, and $\rho_0(\mathbf{r})$ are the acoustic particle velocity, SOS, acoustic density, and ambient density, respectively.
The functions $\tau(\mathbf{r}) = -2 \alpha(\mathbf{r})c^{y-1}(\mathbf{r})$ and $\eta(\mathbf{r})=2 \alpha(\mathbf{r})c^{y}(\mathbf{r})\text{tan}(\pi y/2)$ are the absorption and dispersion proportionality coefficients where $\alpha(\mathbf{r})$ and $y$ are the AA and power law exponent, respectively.
The selection of the transducer for wave emission can be conducted in a sequential manner, starting with the first one and moving on to the subsequent ones.
Let $p_m(\mathbf{r},t)$ and $s_m(\mathbf{r},t)$ denote the pressure wavefield and source function at $m$-th emitter with $m = 0, ..., N_e-1$ for $N_e$ denoting the total number of emitters.
By introducing a shorthand notation for Equations \eqref{eq:waveeqs}, \eqref{eq:waveeqs2}, and \eqref{eq:waveeqs3}, the wavefield $p_m(\mathbf{r},t)$ can be represented in an operator form:
\begin{equation}
    p_m(\mathbf{r}, t) = \mathcal{H}^{\mathbf{a}} s_m(\mathbf{r}, t),
\end{equation}
where the linear operator $\mathcal{H}^{\mathbf{a}}:\mathbb{L}^2(\mathbb{R}^3 \times [ 0,T))\rightarrow \mathbb{L}^2(\mathbb{R}^3 \times [ 0,T))$ describes the action of the wave equation, which explicitly depends on the functions of acoustic properties $\mathbf{a} = \left[c,\rho,\alpha,y \right]$.
For all receivers $n=0, 1, ..., N_r - 1$ with $N_r$ denoting the total number of receivers located on the continuous measurement aperture $\Omega\subset \mathbb{R}^3$, the measurement $g_{m,n}(t)\in\mathbb{L}^2([0,T))$ for the $n$-th receiver located at $\mathbf{r}_n\in\Omega$. 
This measurement can be described using the imaging model as a continuous-to-continuous (C-C) mapping as:
\begin{equation}\label{eq:c-c}
    g_{mn}(t) = \mathcal{M}_n \mathcal{H}^{\mathbf{a}} s_m(\mathbf{r}, t).
\end{equation}
Here, the operator $\mathcal{M}_n:\mathbb{L}^2(\mathbb{R}^3 \times [ 0,T))\rightarrow \mathbb{L}^2([0, T))$ is the sampling operator to map the measurement corresponding to the $n$-th receiving transducer.
For a point-like receiver, the measurement $g_{mn}(t)$ can be represented as $\int_{\mathbb{R}^3}\delta(\mathbf{r}-\mathbf{r}_n) \mathcal{H}^{\mathbf{a}} s_m(\mathbf{r}, t) d\mathbf{r}$.

Given the discrete sampling effects in a digital USCT imaging system, the C-C mapping presented in Eq. (\ref{eq:c-c}) can be represented by a continuous-to-discrete (C-D) mapping.
The finite-dimensional representation of the functions $g_{mn}(t)$, denoted as $\mathbf{g}_{mn}$, is obtained by sampling them at temporal interval $\Delta t = T/L$, where $L$ denotes the number of time samples. 
This can be expressed as:
\begin{equation*}
    \left[ \mathbf{g}_{mn}\right]_l \equiv g_{mn}(l\Delta t),
\end{equation*}
for $l=0,1, ..., L-1$.
In summary, given these discretized quantities, the C-D mapping model for USCT can be described as:
\begin{equation}
     \mathbf{g}_{mn}= \mathcal{S} \mathcal{M}_n\mathcal{H}^{\mathbf{a}}s_m(\mathbf{r},t),
\end{equation}
where $\mathcal{S}: \mathbb{L}^2([0,T)) \rightarrow \mathbb{R}^L$ denotes the temporal sampling operator.

\subsection{Bent ray-based traveltime tomography (BRTT)}
\label{sec:backgrounds:ray}
BRTT methods are one efficient way of estimating the slowness distribution of tissues from USCT data \cite{hormati2010robust, quan2007sound,norton1982correcting,ali2019open}.
Here, the slowness distribution, denoted as $\mathbf{b}\in\mathbb{R}^K$, is a finite-dimensional representation of the reciprocal of the SOS distribution $c(\mathbf{r})$, defined as $\left[ \mathbf{b}\right]_k = 1/c(\mathbf{r}_k)$ for $k=0,1,..., K-1$, where $\mathbf{r}_k$ denotes the $k$-th spatial grid point.

These methods employ the time-of-flight (TOF) of ultrasound waves transmitted through tissues to reconstruct the slowness distribution.
Let $\mathbf{t}_m\in\mathbb{R}^{N_r}$ denote the predicted TOF vector corresponding to the $m$-th emitter, where its $n$-th element $\left[\mathbf{t}_m\right]_n$ corresponds to the TOF measured at the $n$-th receiver. 
Using geometrical acoustics, the TOF can be interpreted as a line integral through the slowness distribution along the  ray-path of ultrasound propagation \cite{ali2019open}.
Specifically, $\mathbf{t}_m$ for the slowness $\mathbf{b}$ can be expressed as $\mathbf{t}_{m} = \mathbf{L}^{\mathbf{b}}_m \mathbf{b}$, where $\mathbf{L}^{\mathbf{b}}_m\in\mathbb{R}^{N_r\times K}$ is the so-called ray-tracing matrix representing the line integrals performed over the slowness $\mathbf{b}$.
Here, the $(n,k)$-th element, $\left[\mathbf{L}^{\mathbf{b}}_m\right]_{n,k}$, represents the arclength of the ray-path between emitter $m$ and receiver $n$ within the $k$-th pixel \cite{ali2019open}.

A common BRTT approach seeks an estimate of the slowness distribution vector, denoted as $\hat{\mathbf{b}}\in \mathbb{R}^K$, by minimizing the sum of squared differences between the observed TOF and the predicted one for all $(m,n)$-pairs of the emitters and receivers.
Let $\mathbf{t}^{obs}_{m}$ denote the observed TOF corresponding to the $m$-th emitter, where the $n$-th element, $\left[\mathbf{t}^{obs}_{m}\right]_n$, is obtained by measuring the first arrival based on the measured data $\mathbf{g}_{mn}$.
The estimate $\hat{\mathbf{b}}$ can be obtained by solving the optimization problem:
\begin{equation}
\label{eq:backgrounds:ns}
\hat{\mathbf{b}} = \argmin_{\mathbf{b}\in \mathbb{R}^K}\sum_{m=0}^{N_e-1} \|\mathbf{t}^{obs}_m -\mathbf{L}^{\mathbf{b}}_m \mathbf{b} \|_2^2.
\end{equation}
The corresponding estimated SOS distribution vector, denoted as  $\mathbf{c}_{Low}\in \mathbb{R}^K$, is obtained as:
\begin{equation}
\left[ \mathbf{c}_{Low} \right]_k = 1/\left[\hat{\mathbf{b}}\right]_k,
\end{equation}
for $k=0,1,...,K-1$. 
Here, the subscript \textit{Low} indicates that the estimated SOS possesses low spatial resolution.
This optimization problem is typically solved by the Gauss-Newton (GN) method because it is a well-established method for solving nonlinear least-squares problems \cite{ali2019open}.  
In the studies presented below, a box constraint is imposed on the slowness distribution perturbation, which involves the use of an indicator function as a non-smooth regularization term.
This modification requires the use of a proximal method for computing the GN update.
Mathematical details and an algorithm for this approach are described in Appendix \ref{app}.

\subsection{Delay-and-sum-based reflection tomography (DAS-RT)}
\label{sec:backgrounds:das}
The DAS imaging technique is widely used for RT \cite{stotzka2002medical,fitch2012synthetic,kretzek2014gpu,duric2015whole,qu2018effect}. 
A typical approach involves delaying the received back-scattered signals based on the assumed SOS and then summing them to form an image of the reflectivity distribution.
This summing process is typically performed over multiple emitter-receiver pairs to improve the signal-to-noise ratio (SNR).
The discretized DAS image, denoted as a vector $\mathbf{f}\in\mathbb{R}^K$, can be represented as:
\begin{equation}
\label{eq:backgrounds:das}
\left[\mathbf{f}\right]_k = \sum_{m=0}^{N_{e}-1}\sum_{n=0}^{N_{r}-1} a_{m, n} \left[\mathbf{g}_{mn}\right]_{l_k}.
\end{equation}
Here, the index $l_k$ denotes a discrete time-step; specifically, the quantity $l_k\Delta t$ represents the two-way traveltime of a signal, originating from the $m$-th transmitter, reflecting at the $k$-th grid point, and subsequently being received by the $n$-th receiver
The quantity $a_{m,n}$ denotes an apodization value, and it is specifically designed to account for only the back-scattered signals as follows:
\begin{align}
a_{m,n} =
\begin{cases}
1 &\text{ if } \angle_{m,n} \le \sigma\ \\
0 &\text{ otherwise, } \
\end{cases}
\end{align}
where $\angle_{m,n}$ represents the central angle subtending the arc between the $n$-th receiver and the $m$-th emitter, and $\sigma$ is a threshold value.

\section{Methods}
\label{sec:methods}
The studies below seek to understand the benefits of employing a low-resolution SOS map produced by BRTT and a reflectivity map produced by DAS-RT as concurrent inputs for a learned image-to-image reconstruction operator for USCT  based on a CNN. 
The target image (i.e., desired output of the CNN) is a high-resolution SOS map.
Here, the low-resolution SOS map serves as an initial, coarse approximation of the SOS distribution in the tissue, retaining low spatial frequency components while offering direct quantitative information. 
The reflectivity map, on the other hand, provides high-spatial-frequency information about the desired SOS associated with tissue boundaries.
This map is particularly sensitive to contrast variations in the desired SOS map through acoustic impedance differences across tissues. 
The subsequent sections describe this dual-channel CNN-based reconstruction method as well as a fine-tuning approach to improve tumor reconstruction accuracy.

\subsection{Dual-channel CNN-based reconstruction via a U-Net}
U-Net-based approaches have shown promising performance in a wide range of medical image reconstruction tasks \cite{ronneberger2015u, jin2017deep, ghodrati2019mr}.
The U-Net utilizes skip-connections to combine features from the contracting path with those in the expanding path, which improves the localization accuracy of the network. 
Moreover, the U-Net employs pooling layers to extract higher-level features from the input image, which reduces the dimensionality of the feature maps and helps to mitigate overfitting when training on limited datasets.
Additionally, the U-Net can be adapted to handle any number of input channels, making it well-suited for the proposed dual modality approach. 

In the context of SOS reconstruction, the U-Net approach seeks to minimize the mean squared error (MSE) between the predicted and target SOS images.
For the 2D reconstruction problem, the target SOS vector $\mathbf{c}\in\mathbb{R}^K$, the SOS vector reconstructed by the BRTT $\mathbf{c}_{Low}\in\mathbb{R}^K$, and the reflectivity vector reconstructed by the DAS-RT, denoted as $\mathbf{f}\in\mathbb{R}^K$, are reshaped into matrices as follow:
\begin{equation*}
    \left[\mathbf{C}\right]_{j,l} = c(\mathbf{r}_k),      \left[\mathbf{C}_{Low}\right]_{j,l} = \left[\mathbf{c}_{Low}\right]_k,\text{ and } \left[\mathbf{F}\right]_{j,l} = \left[\mathbf{f}\right]_k,
\end{equation*}
with $k = N_x\cdot l + j$ for $j=0,1,...,N_x-1$ and $l = 0,1,...,N_y-1$. Here, the quantities $N_x$ and $N_y$ denote the sizes of each dimension of a 2D image.
The input to the U-Net is a tensor represented by the dual-channel $(\mathbf{C}_{Low}^i, \mathbf{F}^i) \in \mathbb{R}^{2\times N_x \times N_y}$. 
The index $i$ spans the range $0 \leq i < I$, indicating the training sample number.
Given a U-Net-based reconstruction operator $\mathcal{A}_{\boldsymbol{theta}}:\mathbb{R}^{2\times N_x \times N_y}\rightarrow \mathbb{R}^{N_x\times N_y}$ that is 
parameterized by the neural network weights  $\boldsymbol{\theta}\in\mathbb{R}^{W}$, the standard MSE loss is defined as:
\begin{equation} 
L(\boldsymbol{\theta}) =  \frac{1}{2I}\sum_{i=1}^I ||\mathcal{A}_{\boldsymbol{\theta}}(\mathbf{C}_{Low}^i, \mathbf{F}^i) - \mathbf{C}^i||^2_F, \end{equation}
where $\| \cdot \|_F$ denotes the Frobenious norm of the matrix (i.e. the sum of squared entries).

In USCT, the SOS within the water bath is assumed to be constant for each object but can vary across an ensemble of objects due to multiple factors that include temperature variations \cite{marczak1997water}.
Variability in the water bath SOS value within the training dataset can increase the difficulty of the learning problem. 
To address this, for use in training the U-Net, the input SOS reconstructed by the BRTT, $\mathbf{C}_{Low}$, and the target SOS maps, $\mathbf{C}$, are replaced with new quantities in which the corresponding water bath SOS values are subtracted out.
Specifically, those two maps are modified as $\underline{\mathbf{C}}_{Low}^i =\mathbf{C}_{Low}^i- \mathbf{C}^i_{w}$ and $\underline{\mathbf{C}}^i = \mathbf{C}^i - \mathbf{C}_{w}^i$, where $\mathbf{C}^i_{w}$ represents the matrix whose elements are equal to the constant water bath SOS value of the $i$-th training sample. 

In terms of these quantities, the standard MSE loss function can be expressed as:
\begin{equation} 
L(\boldsymbol{\theta}) =  \frac{1}{2I}\sum_{i=1}^I ||\mathcal{A}_{\boldsymbol{\theta}}(\underline{\mathbf{C}}_{Low}^i, \mathbf{F}^i) - \underline{\mathbf{C}}^i||^2_F. \end{equation}

The U-Net is trained  by seeking approximate solutions to the following non-convex optimization problem:
\begin{equation} 
\boldsymbol{\theta}^{MSE} = \argmin_{\boldsymbol{\theta} \in \mathbb{R}^W} L(\boldsymbol{\theta}),\end{equation}
where the superscript $MSE$ denotes that the employed training loss is the standard MSE loss.
During training, the network learns to map the dual-channel input tensor to a high-quality image that closely matches the target SOS map. 
This is achieved by minimizing the standard MSE loss between the predicted and target images using an optimizer such as stochastic gradient descent.
This dual-channel U-Net is denoted as U-Net-RT throughout this paper, where R and T denote reflectivity and traveltime tomography images, respectively, which are used as inputs.

After the U-Net-RT is trained, a high-resolution SOS estimate $\bar{\mathbf{C}}$  can be computed from a newly acquired pair $(\underline{\mathbf{C}}_{Low}, \mathbf{F})$ of BRTT and DAS-RT images as
\begin{equation}
    \bar{\mathbf{C}} = \mathcal{A}_{\boldsymbol{\theta}^{MSE}}(\underline{\mathbf{C}}_{Low}, \mathbf{F}) + \mathbf{C}_w,
\end{equation}
where the matrix $\mathbf{C}_w$ has elements equal to the constant water SOS value of $\mathbf{C}_{Low}$, which is known.

\subsection{Tumor-weighted fine-tuning}\label{met:fine}
The learned reconstruction method described above employs the standard MSE loss for U-Net-RT training and therefore does not preferentially weight image features that may be diagnostically relevant. 
In cases of imbalanced data where certain tissue types are underrepresented in the training data, the reconstruction accuracy for such tissues can be degraded.
To mitigate this issue, a fine-tuning strategy is proposed to refine the base network, which is initially trained by use of the standard MSE loss function.

Specifically, a fine-tuning strategy is proposed to refine the base network, which is initially trained by use of the standard MSE loss function.
Specifically, in this fine-tuning process, the MSE loss function is adjusted to give higher weights to the minority class of tissue types. 
This weighting ensures that when computing the MSE loss, the U-Net-RT model becomes more sensitive to these underrepresented tissues.
This fine-tuning approach becomes especially important when the reconstructed images are to be used for detecting lesions whose prevalence in the training data is low. 

Below, this approach is implemented as follows. Given a weight matrix $\mathbf{W}\in \mathbb{R}^{N_x \times N_y}$ that assigns different values to various tissue types according to the target SOS image, a weighted mean squared error (WMSE) loss function is defined:
\begin{equation} \label{tumor-weighted loss}
L_{w}(\boldsymbol{\theta}) =  \frac{1}{2I}\sum_{i=1}^I ||\mathbf{W}\odot\left(\mathcal{A}_{\boldsymbol{\theta}}(\underline{\mathbf{C}}_{Low}^i, \mathbf{F}^i) - \underline{\mathbf{C}}^i\right)||_F^2, \end{equation}
where the operator $\odot$ indicates an element-wise product. 
In the fine-tuning process, the model weights $\boldsymbol{\theta}$ are initialized as $\boldsymbol{\theta}^{MSE}$ obtained by pre-training the U-Net-RT by use of the standard MSE loss.
Subsequently, the weights are further refined by finding approximate solutions of the following minimization problem:
\begin{equation} 
\boldsymbol{\theta}^{WMSE} = \argmin_{\boldsymbol{\theta} \in \mathbb{R}^W} L_{w}(\boldsymbol{\theta}). \end{equation}

To facilitate the weighting of tumor regions, the elements of $\mathbf{W}$ can be assigned as:
\begin{align*}
    \left[\mathbf{W}\right]_{j,l} = \begin{cases}
        0& \text{ if } \mathbf{r}_k \text{ is located in the water-bath} \\
        w& \text{ else if } \mathbf{r}_k \text{ is located in tumor regions} \\
        1& \text{ otherwise}
    \end{cases} 
\end{align*}
with $k = N_x\cdot l + j$ for $j=0,1,...,N_x-1$ and $l=0,1,...,N_y-1$.
Here, the weight value $w$ is set to be greater than $1$.
Likewise, the SOS estimate produced by the fine-tuned U-Net-RT model can be obtained as $\mathcal{A}_{\boldsymbol{\theta}^{WMSE}}(\underline{\mathbf{C}}_{Low}, \mathbf{F}) +\mathbf{C}_w$.

\section{Virtual imaging Studies}
\label{sec:numerical studies}
Virtual imaging studies were designed to systematically investigate the use of traveltime tomography and reflection tomography as input modalities for IILR-based SOS estimation in USCT based on virtual imaging studies.
These inputs correspond to low-resolution SOS maps produced by BRTT and reflectivity maps produced by DAS-RT. 
The IILR method was implemented using the U-Net architecture. 
An evaluation of the generalizability of the dual-channel U-Net was also conducted. 
Both traditional U-Net-based reconstruction methods and fine-tuning approaches were objectively assessed by use of a binary signal detection task. 
Further details of these virtual studies are provided below.

\subsection{Generation of phantoms}
\label{sec:ns:phantom}
The USCT data were produced by use of anatomically realistic numerical breast phantoms (NBPs) \cite{badano2018evaluation, li20213}. 
A tumor-free NBP was defined by 3D maps that describe the SOS, AA, and density distributions with consideration of ten different tissue types that include fat, glandular tissue, ligaments, and skin.

Additionally, as described in Li \textit{et al.} \cite{li20213}, these phantoms incorporate within-tissue acoustic heterogeneity through the addition of spatially correlated Gaussian random field perturbations to the SOS and density distributions \cite{franceschini20062}. 

Based on breast density, a phantom was assigned one of the four BI-RADS classifications: A) almost entirely fatty breasts, B) breasts with scattered fibroglandular density, C) heterogeneously dense breasts, and D) extremely dense breasts \cite{li20213, bi-rads}.
Large ensembles of these healthy NBPs were produced by use of previously developed computational tools \cite{badano2018evaluation, li20213}. 
The specifics of the object variations, such as size distributions, texture patterns, and acoustic properties, were determined using the same methodologies described in Li \textit{et al}. \cite{li20213}.

Similarly, a large collection of numerical tumor phantoms was also generated. 
The features of these tumors including size, shape, and locations within the breast were determined based on probabilistic models as referenced from \cite{li20213}. 
Subsequently, these tumor phantoms were inserted into tumor-free NBPs to create phantoms with tumors.
Finally, 2D cross-sectional slices capturing SOS, AA, and density were extracted at a specific elevation. 
These 2D SOS, AA, and density maps were employed as objects to simulate 2D USCT data.

\subsection{Virtual USCT breast imaging system}
\label{sec:numerical studies:imagingsystem}

The virtual imaging system consisted of a circular measurement aperture of radius 110 mm, which is representative of existing ring-array USCT breast imagers \cite{duric2014clinical, roberts2021transducer}. Along the aperture, 256 point-like (\emph{idealized}) transducers were evenly arranged, each operating as both emitter and receiver.
The central frequency of the source pulse was set to 1 MHz, and the acquisition time for each view was 170 $\mu$s.

 USCT measurement data were simulated by numerically solving a 2D wave equation for an acoustically heterogeneous and attenuating medium. Specifically, the lossy acoustic wave equation in the first-order system formulation (Eqs. \eqref{eq:waveeqs}, \eqref{eq:waveeqs2}, and \eqref{eq:waveeqs3}) was solved using the pseudospectral \textit{k}-space method \cite{treeby2010photoacoustic}. 
This method efficiently solves the wave equation by using Fourier transforms to compute spatial derivatives, enabling accurate modeling of wave propagation in complex media while reducing computational cost. 
A computational grid with 2560-by-2560 pixels, each sized at 0.1 mm, was employed in the simulation.
The timestep was set to 0.02 $\mu$s (Courant-Friedrichs-Lewy number = 0.32 for maximum SOS of 1.6 mm/$\mu$s).
To minimize boundary reflections due to the finite size of the computational domain, perfectly matched layers (PMLs) \cite{berenger1994perfectly} with a thickness of 4 mm at each edge of the grid were implemented. 
The simulation parameters of the 2D virtual imaging system are reported in Table \ref{tab:sim_param}.
Finally, the simulated measurements were corrupted with independent identically distributed Gaussian noise with zero mean.
The noise level was set such that the signal-to-noise ratio (SNR) was 36 dB at the receiver located diagonally opposite the emitter.
Here, the SNR represents the logarithmic ratio of a noiseless signal's power to the noise power.

\begin{table}[!t]
\renewcommand{\arraystretch}{1.3}
\caption{Simulation parameters of the 2D virtual imaging system}
\label{tab:sim_param}
\centering
\begin{tabular}{lll}
\hline

Number of emitters/receivers & 256/256 \\
Radius of ring-array & 110 \textit{mm} \\
Central frequency of source pulse & 1 \textit{MHz} \\
Computational grid & [2560, 2560] \\
Pixel size & 0.1 \textit{mm} \\ 
Acquisition time for each view & 170 $\mu s$ \\
Timestep & 0.02 $\mu s$ \\
Transducers' property & Point-like \\
PML thickness & 4 mm \\
\hline
\end{tabular}
\end{table}

\subsection{Generation of training data}\label{ns:training data}
Using the virtual imaging framework described above, training, validation and testing datasets were created for training and assessing the U-Net-based reconstruction methods. A set of 1,300 unique 3D NBPs was generated as described in \cite{li20213} and a single 2D cross-sectional slice was extracted from each 3D NBP and virtually imaged as described in Section \ref{sec:ns:phantom}. A representative subset consisting of 52 target SOS maps and corresponding USCT simulated measurement data is available on Harvard Dataverse \cite{DVN/CUFVKE_2021}. Below, the procedures for generating input and target images for the training of the U-Net-based reconstruction methods are described.
\subsubsection{True SOS map (Target)}
The target image sets for the network were defined using the 2D SOS maps extracted from the NBPs.
The SOS maps were subsequently resized to dimensions of 1024-by-1024 pixels with a pixel size of 0.25 mm, utilizing nearest-neighbor (NN) interpolation. 
The resulting target SOS maps were cropped to 592-by-592 pixels, removing the uniform water bath region while ensuring the entire breast object is included without truncation.

\subsubsection{Low-resolution SOS map (Input)}\label{ns:bent}
To generate low-resolution SOS maps, the BRTT method was employed, as detailed in Section \ref{sec:backgrounds:ray}.
The TOF was estimated for all emitter-receiver pairs using the Akaike Information Criterion (AIC) picker method \cite{li2009improved}.
To apply the AIC picker, a window of [$T-6\mu s$, $T+6\mu s$] was set for each emitter and receiver pair, where $T$ is TOF of the pulse between the emitter and receiver in a homogeneous medium (water bath).
Before applying the AIC picker, a 4th-order Butterworth low-pass filter with a cut-off frequency of 1.5 MHz was applied to the measurement data to reduce noise.

The algorithmic details for SOS reconstruction using the proximal GN method with a box constraint are described in Algorithm \ref{alg:pgn-box} in Appendix \ref{app}. 
The reconstruction was performed on a 256-by-256 grid with a pixel size of 1mm, and the transducer locations were mapped onto the same grid using NN interpolation.
The reciprocal of the reconstructed slowness image was computed to obtain the SOS estimate.
The resulting low-resolution SOS map was upscaled using NN interpolation to 1024 by 1024 grid with a 0.25mm pixel size and then cropped to 592 by 592 pixels.
 
\subsubsection{Reflectivity map (Input)}
For reconstructing high-resolution reflectivity maps, the DAS-based beamforming method was employed, as described in Section \ref{sec:backgrounds:das}. 
The two-way travel time was computed by solving the 2D Eikonal equation, utilizing the slowness model obtained from BRTT.  
A threshold value $\sigma = \frac{15\pi}{128}\text{rad}$ was used, leading to the inclusion of back-scattered signals from 15 adjacent receivers (7 on each side) for each emitter.
The grid points were located on a 1024-by-1024 grid with a pixel size of 0.25mm, and the transducer locations were mapped onto the same grid using NN interpolation. 
To improve the spatial resolution of the reflectivity map, the measured signals $\underline{\mathbf{g}}_m$ were deconvolved with the source pulse before interpolating them to obtain $t_{m,n}$ \cite{lasaygues2006assessing}. 
Subsequently, the deconvolved $\underline{\mathbf{g}}_m$ were interpolated onto the two-way traveltime $t_{m,n}$ using modified Akima piecewise cubic Hermite interpolation \cite{akima1970new, akima1974method}. 
Finally, the reflectivity images were obtained by summing the interpolated back-projected signals over all emitters and then cropped to 592-by-592 pixels.

\subsection{Training and validation of the dual-channel U-Net}\label{train}
\subsubsection{U-Net architecture}
The contracting path was composed of six blocks, each consisting of two 3$\times$3 convolutional layers followed by a 2$\times$2 average pooling layer. 
The number of output channels for the first block was set to 32, and it was doubled at each subsequent block until reaching 1024 channels.
The expanding path mirrored the contracting path and was composed of six blocks, each consisting of a 2$\times$2 transposed convolutional layer followed by two 3$\times$3 convolutional layers. Batch normalization was incorporated prior to each convolutional layer \cite{ioffe2015batch}. 
The activation function used in all convolutional layers except the last one in the expanding path was a leaky ReLU with a negative slope of 0.2 \cite{maas2013rectifier}.
There was no activation function applied to the output layer. 
The U-Net model comprised a total of 31.1 million trainable parameters.

\subsubsection{Dataset and U-Net training details} 
To train the traditional U-Net model using the standard MSE loss, the dataset was split into training, validation, and testing sets. 
The training, validation, and testing sets consisted of 1,120, 90, and 90 examples of  2D cross-sectional BRTT and DAS-RT images (network inputs) and the corresponding target SOS (i.e., true SOS), respectively. 
The optimization method used was Adam \cite{kingma2014adam} with cycling learning rate \cite{smith2017cyclical}, where the learning rate was linearly varied between 1e-2 and 1e-5 over a period of 2000 iterations. 
The batch size was 80, and the training was performed on 4 NVIDIA Tesla V100 PCIe 32GB GPUs. 
To prevent overfitting and improve the generalization capability of the U-Net-based methods, data augmentation was employed with a non-augmented to augmented data ratio of 1:3 using random rotations and horizontal and vertical flips at each epoch. 
The U-Net model was trained for a maximum of 5,000 epochs and the best model was the one that achieved the smallest validation loss during those epochs.

\subsection{Study designs}
\subsubsection{Study 1 - Investigating the impact of dual-modality inputs}\label{num:study1}
The first study aimed to understand the impact of the use of dual-channel inputs on high-resolution SOS reconstruction using the U-Net-based method. 
For this purpose, the dual-channel U-Net-RT was trained using the standard MSE loss function, following the training procedure detailed in Section \ref{train}.
The SOS maps produced by U-Net-RT were qualitatively evaluated on the testing dataset.

The images produced by use of the U-Net-RT were compared both qualitatively and quantitatively to those produced by two separate single-channel U-Nets: U-Net-T, trained on low-resolution SOS maps generated by BRTT, and U-Net-R, trained on reflectivity maps from DAS-RT.
This comparison aimed to understand the contribution of each input modality to the network's performance.
Although these models shared the same network architecture as U-Net-RT, they differed in terms of their number of input layers: the U-Net-RT utilized two channels, whereas U-Net-T and U-Net-R used only one. 
The U-Net-R and U-Net-T models were trained by use of the same procedures employed for U-Net-RT training.

The performance of the U-Net models was compared to that achieved by a 2D time-domain FWI method, which served as a baseline reference \cite{wang2015waveform}.
This FWI-based reconstruction method was implemented using source encoding, following the approach described in Wang \textit{et al}. \cite{wang2015waveform}.
The  FWI method was initialized using the low-resolution SOS estimate produced by BRTT, as described in Section \ref{ns:bent}.
Note that this SOS map also served as an input for both the U-Net-RT and U-Net-T. 
During the FWI iterative process, both the density and AA coefficient maps were assumed to be known and fixed.  Althought this is an idealization, this assumption is only expected to enhance the performance of the reference FWI method.
The computational grid consisted of 1024-by-1024 pixels with a pixel measuring 0.25mm, which is the same resolution of the output produced by U-Net-RT, U-Net-R, and U-Net-T.
\subsubsection{Study 2 - Evaluation of generalization performance}
The second study aimed to evaluate the generalizability of the dual-channel approach to out-of-distribution (OOD) data.
These data differ significantly from the training data,  and can potentially compromise the performance of the learned reconstruction method. 
In this study, the OOD data corresponding to a specific breast density type that was excluded from the training set.
Specifically, a U-Net-RT model was trained using breast types A, B, and C (1,010 NBPs).
This model, along with the previously described U-Net-RT model that was trained using all breast types (1,120 NBPs) from Section \ref{num:study1},
were evaluated using test data corresponding to 18 breast type D NBPs. It should be noted that, on average, breast type D   is smaller in size and has stronger acoustic heterogeneity than breast types A, B and C.  As such, breast type D can be interpreted as
 OOD data for the first U-Net-RT model mentioned above.

\subsubsection{Study 3 - Impact of rare tumor occurrence within the training dataset} \label{num:study3}
The third study aimed to quantitatively assess the local reconstruction accuracy within tumor regions as well as task-based performance associated with a tumor detection task using the dual-channel U-Net approach.
It should be noted that tumor tissues accounted for a mere 0.22$\%$ of the total breast tissue area within the training dataset.

\begin{figure}[!h]
    \centering
    \includegraphics[width=0.48\textwidth]{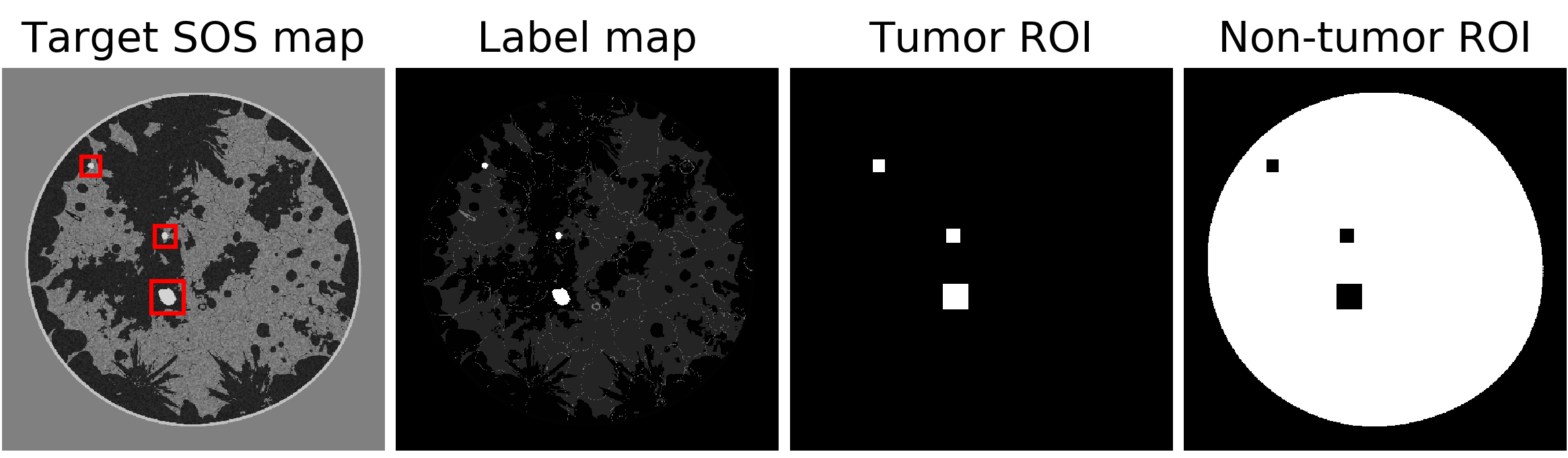}
    \includegraphics[width=0.48\textwidth]{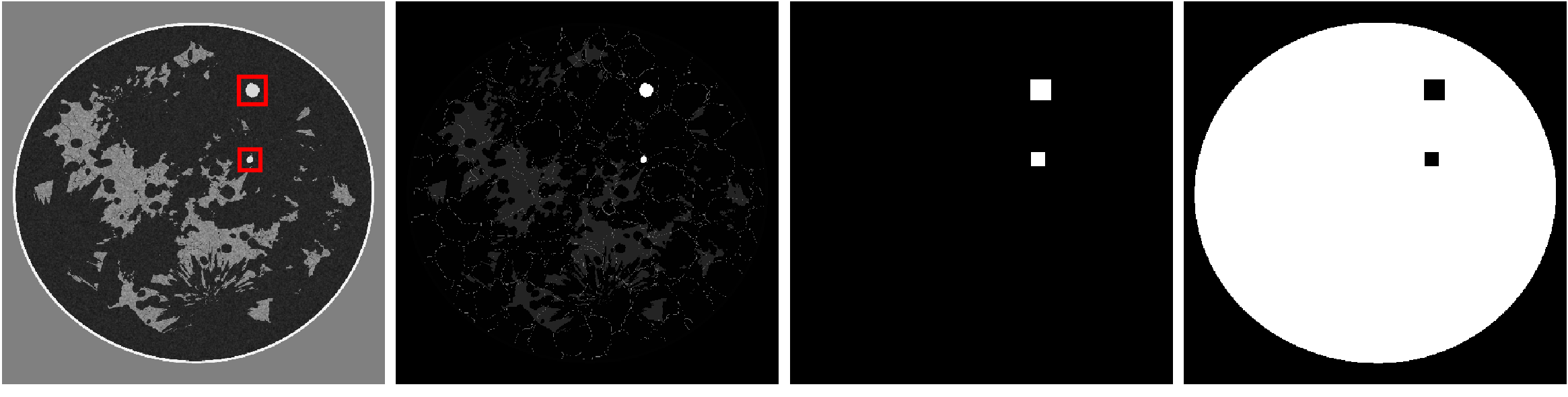}
    \caption{Illustrative examples of tumor and non-tumor regions within breast images. From left to right: target SOS maps displaying tumor ROIs (red boxes), label maps highlighting tumor locations, binary maps of tumor ROIs, and binary maps of normal breast tissue excluding the tumor ROIs. Both tumor and normal tissue regions were employed for the calculation of traditional IQ metrics, as well as for determining distinct weight values for each region when fine-tuning the networks using the WMSE loss.}
    \label{fig:label}
\end{figure}

Reconstruction accuracy for both tumor and non-tumor regions was assessed in the SOS estimates yielded by the U-Net-RT, U-Net-R, U-Net-T, and FWI.
The implementations of these four reconstruction methods were detailed in Section \ref{num:study1}.
To facilitate this assessment,  regions of interest (ROIs) of tumor tissues were identified by using segmented label maps of the corresponding target SOS maps, where each label denotes a specific tissue type, including tumor. 
Based on the label maps, the width and height of each tumor were determined. 
The larger of these two dimensions was selected as the side length of a square ROI, which was then centered around the tumor.
Subsequently, this tumor ROI was expanded by 4 pixels on each side. 
It should be noted that the size of the  ROI could vary between different tumors.
Examples of the tumor ROIs and normal tissues region along with the target SOS and label maps are displayed in Fig. \ref{fig:label}.

The U-Net-RT model, initially trained using the standard MSE loss, was subjected to the fine-tuning process using the WMSE loss function, as detailed in Section \ref{met:fine}.
To assign loss function weights to tumor areas and normal tissue regions, the tumor and non-tumor ROIs  described above were employed.
The training data contained 1,023 tumor-containing samples out of a total of 1,120, while the testing set included 82 tumor-containing samples out of a total of 90. 
Each tumor-included case could contain one to three isolated tumors.
To investigate the impact of the weight value $w$ of the WMSE loss function, described in Eq. \eqref{tumor-weighted loss}, multiple instances of the pre-trained U-Net-RT model were fine-tuned using a distinct $w$ value for each instance.
Specifically, four different instances of U-Net-RT were fine-tuned corresponding to different values of $w=2,5,10,20$.
Hereafter, these differently fine-tuned U-Net-RT models will be denoted as U-Net-RTw2, U-Net-RTw5, U-Net-RTw10, and U-Net-RTw20.
Each fine-tuning process closely followed the training methodology used for training the standard U-Net-RT but with minor modifications in the maximum cyclic learning rate, which was decreased from 1e-2 to 1e-3, and the maximum epoch set at 2,000.
The fine-tuned models with the lowest WMSE loss on the validation set were selected.

To assess tumor detection performance, this study involved a patch-based signal-known-stochastically (SKS) and background-known-stochastically (BKS) binary detection task\cite{barrett2013foundations, li2023estimating}. 
This task involved testing a hypothesis with two alternatives: either at least one tumor is present within a patch in the reconstructed image, or the signal is absent within the patch.
This binary signal detection task requires an observer to predict the true hypothesis given a reconstructed SOS map.
The performance of a numerical observer (described below) was objectively assessed using reconstructed SOS maps for each reconstruction method including the U-Net-RT, U-Net-R, U-Net-T, fine-tuned U-Net-RT models, and FWI.
The details regarding this procedure are provided in the following subsection.

\subsection{Performance evaluation} \label{performance}
\subsubsection{Traditional IQ measures}
The performances of the U-Net-based reconstruction methods were assessed using three traditional IQ metrics: the ensemble average of the normalized root mean squared error (NRMSE), structural similarity index measure (SSIM) \cite{wang2004image}, and peak signal-to-noise ratio (PSNR).
The NRMSE was used to measure the difference between the target and reconstructed SOS maps, normalized by the difference between the true object and a constant SOS value in the water bath.
Specifically, this metric was calculated as follows:
\begin{equation}
NRMSE = \frac{\|\mathbf{C} - \bar{\mathbf{C}}\|_F}{\|\mathbf{C} - \mathbf{C}_w\|_F}.
\end{equation}
Here, as a reminder, the matrices $\mathbf{C}$, $\bar{\mathbf{C}}$, and $\mathbf{C}_{w}$ denote the target SOS map, the SOS estimate produced by the U-Net-based reconstruction methods, and the matrix with elements of the constant SOS value in water $c_w$, respectively.
Given that the dynamic range of all SOS maps is similar in this study, from 1.4 to 1.6 mm/$\mu$s, the NRMSE is approximately proportional to the root mean squared error.
The SSIM between each target and reconstructed SOS map pair was then computed using the default parameters, as suggested in \textit{Wang et al. } \cite{wang2004image}.
The PSNR was calculated after linearly scaling the image values from their original dynamic range of [1.4, 1.6] to [0, 1].

The statistical significance of differences in traditional IQ metrics was assessed using a non-parametric Mann-Whitney U test \cite{mann1947test, wilcoxon1992individual}.
Specifically:
\begin{itemize}
    \item In Study 1, the U-Net-RT model was compared with each of the following: the U-Net-R, U-Net-T, and FWI in terms of NRMSE, SSIM, and PSNR.
    \item In Study 2, the U-Net-RT, U-Net-R, and U-Net-T models, each trained on all types of NBPs, were compared with their counterparts trained exclusively on non D-type NBPs, in terms of NRMSE, SSIM, and PSNR.
    \item In Study 3, the U-Net-RT, U-Net-R, U-Net-T, FWI, and fine-tuned U-Net-RT models---including U-Net-RTw2, U-Net-RTw5, U-Net-RTw10, and U-Net-RTw20---were evaluated, comparing NRMSE computed on non-tumor versus tumor regions.
\end{itemize}
A two-tailed test was conducted, and p-values were calculated. Results were considered statistically significant if the p-value was less than or equal to the pre-set significance level of 0.05.

\begin{figure*}[!htb]
    \centering
    \includegraphics[width=0.9\textwidth]{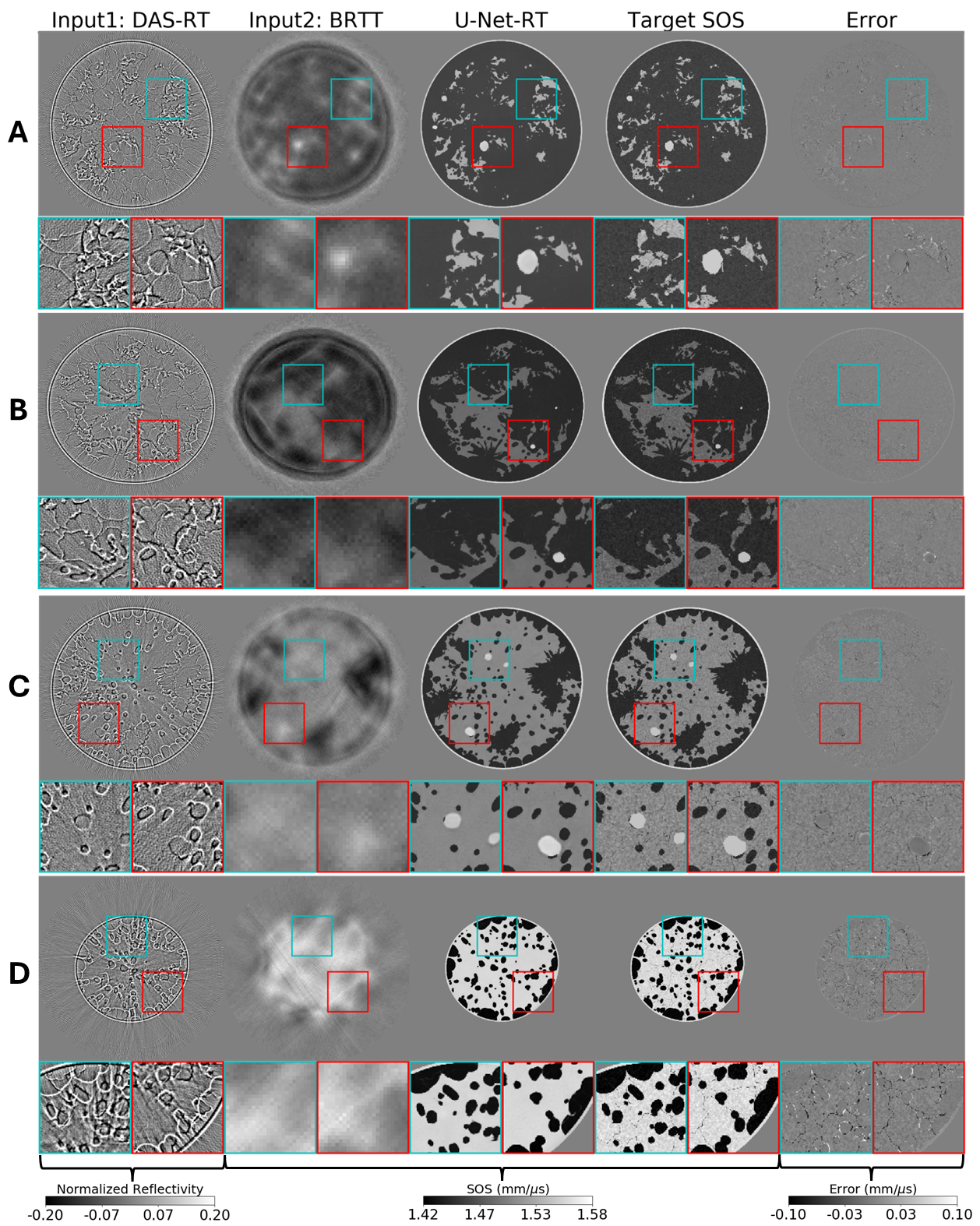}
    \caption{From left to right: Examples of a reflectivity map reconstructed by DAS-RT (scaled using its maximum absolute value), SOS map reconstructed by BRTT, SOS map produced by U-Net-RT, target SOS map, and the corresponding error map representing the difference between the target SOS map and the SOS map produced by U-Net-RT. Images were cropped to a window of size 124mm-by-124mm. From top to bottom: Results corresponding to NBPs representative of the four breast density types (A, B, C, D). Square insets highlight specific patches (25mm-by-25mm) within the images, with zoomed views shown below each image. DAS-RT and BRTT images, serving as inputs for U-Net-RT, contribute tissue boundary information and background SOS information, respectively, to the SOS map produced by U-Net-RT.}
    \label{fig:unet_results_rt}
\end{figure*}

\subsubsection{Task-based IQ assessment}
A numerical observer refers to an algorithm that can perform a specific image-based inference, such as the signal detection task considered here.
In this study, the impact of the input modalities on the U-Net-based reconstruction methods was objectively assessed by computing the performance of a NO performing a patch-based SKS/BKS binary signal detection task, as described in Section \ref{num:study3}.
Specifically, a CNN-based NO was used to compute the posterior probability of signal presence within a given patch to perform the patch-based SKS/BKS binary signal detection task \cite{zhou2019approximating, zhang2021impact}. The design of the CNN-based NO and the task are described as follows.
The CNN-based NO consisted of three convolutional blocks, each with a convolutional, batch normalization, leaky ReLU, and 2 by 2 averaging pooling layers, followed by a fully-connected layer with a sigmoid activation function. 
Each convolutional block had 64 channels and a kernel size of 5 by 5.
This CNN-based observer was pre-trained on target SOS maps within the training dataset, which consisted of 991 images (containing around 1,500 signals) out of a total of 1,120 training samples. 
During training, either a signal-present (SP) or signal-absent (SA) patch, sized 96 by 96 pixels, was randomly extracted from a target image on the fly.
SP patches might include one or more signals positioned randomly within them, possibly including signals from previous epochs but located differently within the new patches.
Subsequently, the pre-trained CNN-based observer was separately fine-tuned using the estimates produced by the relevant reconstructed image type: FWI, U-Net-RT, U-Net-R, U-Net-T, U-Net-RTw2, U-Net-RTw5, U-Net-RTw10, and U-Net-RT20. 
The training dataset for each method consisted of 129 images with 350 unique signals. 
The patch extraction strategy for fine-tuning was the same as the pre-training phase.
For both the pre-training and fine-tuning phases, 5,000 epochs were executed, and the model with the lowest validation loss was selected. 

For the testing phase, a total of 157 SA patches and 155 SP patches were extracted from the testing dataset. 
The evaluation of task-based performance was conducted through receiver operating characteristic (ROC) curve analysis. 
The values of area under the ROC curve (AUC), determined using the respective CNN-based observers, were utilized as the primary figure of merit.
The ROC curves were fit using the Metz-ROC software \cite{metz1998statistical}, employing the binormal model \cite{pesce2007reliable}. 
Furthermore, uncertainties associated with the AUC values were estimated.

\section{Virtual Imaging Results}
\label{sec:results}
\subsection{Study 1 - Investigating the impact of dual-modality inputs}
The impact of dual-modality inputs (TT and RT images) on SOS reconstruction was qualitatively assessed via the U-Net-RT trained using the standard MSE loss.
Examples from the testing dataset are shown in Fig. \ref{fig:unet_results_rt}. 
The input images produced by DAS-RT and BRTT, the SOS maps produced by the U-Net-RT model, the target SOS maps, and the corresponding reconstruction error maps are shown.
The reflectivity images produced by DAS-RT effectively describe the tissue boundaries by capturing the contrast variations resulting from acoustic impedance differences.
The low-resolution SOS maps produced by BRTT effectively captured the overall SOS distribution and provided preliminary background information about the objects.
The use of these dual-channel inputs facilitated the U-Net-RT model in producing accurate and high-resolution SOS map estimates.

\begin{figure}[!htb]
    \centering
    \includegraphics[width=0.505\textwidth]{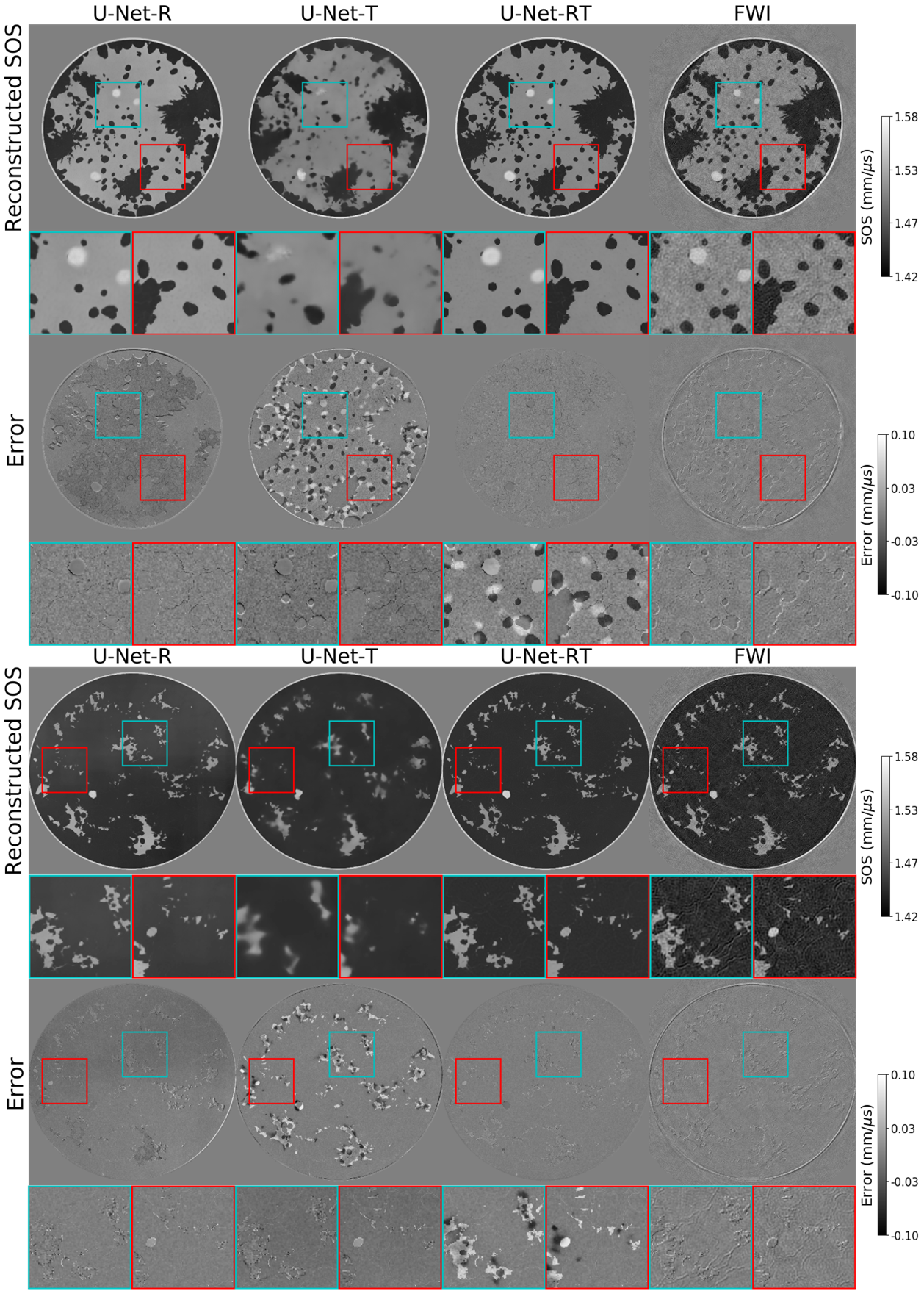}
    \caption{Comparisons of the SOS maps reconstructed by U-Net-R, U-Net-T, U-Net-RT, and FWI (from left to right) and the corresponding error maps, where the examples were drawn from the testing dataset. Images were cropped to a window of size 124mm-by-124mm. Square insets highlight specific patches (25mm-by-25mm) within the images, with zoomed views shown below each image.
    Compared to the U-Net-RT, the estimates from the U-Net-R exhibit bias and those from the U-Net-T contain false structures, while the SOS maps reconstructed by FWI display artifacts.}
    \label{fig:unet_results}
\end{figure}

To gain further insights into the effects of each input modality, a comparative qualitative analysis of the reconstructed SOS maps reconstructed by use of the U-Net-RT, U-Net-R, U-Net-T, and FWI methods, along with their respective error maps, is shown in Fig. \ref{fig:unet_results}. 
It was observed that the SOS estimates produced by the U-Net-R accurately revealed the geometry of the tissue structures. 
However, the corresponding error maps confirmed that the SOS values were inaccurate. This can be explained by the fact that the reflectivity map, which is the input to the U-Net-R method,  does not provide quantitative SOS information to the reconstruction method. 
On the other hand, the U-Net-T effectively reduced bias in the reconstructed SOS estimates by incorporating direct SOS information as input, as evidenced by the corresponding error map. 
However, the reconstructed SOS map was prone to significant artifacts due to the limited high-spatial frequency information in the BRTT input.
This highlights the advantage of utilizing dual modality inputs in the U-Net-RT, as it efficiently combines the strengths of each modality to improve reconstruction results.
Furthermore, the SOS map produced by the U-Net-RT demonstrated a reduced presence of artifacts in comparison to the SOS map produced by FWI.

Table \ref{tab:4all} presents the ensemble-averaged NRMSE, SSIM, and PSNR values, computed on the testing dataset consisting of 18 D-type NBPs.
The U-Net-RT model demonstrated superior performance compared to the U-Net-R, U-Net-T, and FWI models, as established by its lower NRMSE and higher SSIM, and PSNR values.
The statistical significance of the difference between the U-Net-RT and the other approaches was assessed by the use of the two-tailed Mann-Whitney U-test. 
The obtained \textit{p}-values, which are all below 0.01 for the three comparisons, suggest the statistically significant improvement of the U-Net-RT in traditional IQ measures compared to the other approaches.

\begin{table}[!htb]
\renewcommand{\arraystretch}{1.3}
\caption{Study 1: The ensemble average of NRMSE, SSIM, and PSNR for U-Net-RT, U-Net-R, U-Net-T, and FWI evaluated on the testing dataset}
\label{tab:4all}
\centering
\resizebox{\columnwidth}{!}{%
\begin{tabular}{c|c|c|c|c}
\hline
 & \multicolumn{4}{c}{Reconstruction Method} \\
\cline{2-5}
 & U-Net-RT & U-Net-R & U-Net-T & FWI\\
\hline
NRMSE (SD) & 0.2223 (0.0693) & 0.3472 (0.1480) & 0.5414 (0.1491) & 0.3289 (0.0780) \\
\hline
SSIM (SD) & 0.9037 (0.0341) & 0.8848 (0.0405) & 0.8122 (0.0577) & 0.8722 (0.0429) \\\hline
PSNR (SD) & 31.60 (2.489) & 28.04 (3.141) & 23.83 (3.326) & 28.04 (2.030) \\
\hline
\end{tabular}}
\parbox{\linewidth}{\vspace{0.15cm} 
Note: SD stands for standard deviation. PSNR is measured in dB.
}
\end{table}

\subsection{Study 2: Evaluation of generalization performance}\label{sec:results:eval}
To assess the ability of the U-Net-RT to generalize to out-of-distribution data, a comparative analysis was conducted between two variants of U-Net-RT: one trained on all NBP types and another exclusively on non-D type NBPs.
The evaluation of the two U-Net-RT models was conducted on type D NBPs, using the ensemble average of NRMSE, SSIM, and PSNR measures. 
The findings, shown in Table \ref{tab:dtype}, indicate that the U-Net-RT model trained on non-D type NBPs exhibited a high level of accuracy in terms of the NRMSE, SSIM, and PSNR that was comparable to that trained on all types of NBPs when evaluated on type D NBPs, despite not being explicitly trained on this specific NBP type.
The lack of statistical significance between the two models is evident from the estimated \textit{p}-values for NRMSE (0.07), SSIM (0.62), and PSNR (0.21) which did not reach the standard significance threshold of 0.05. 
The findings of this study highlight the U-Net-RT's potential for generalizability since it exhibits satisfactory performance over a wide range of breast types, including those that are significantly underrepresented in the training dataset.
However, it should be noted that, although not shown, when comparing the performance of the two U-Nets on non-D type NBPs, \textit{p}-values larger than 0.8 were observed for the NRMSE, SSIM, and PSNR. 
This implies that while the observed differences between the two U-Net-RT models in relation to type D breasts may not possess statistical significance, there could still be some intrinsic performance gap between the models specifically for this breast type.

\begin{table}[!htb]
\renewcommand{\arraystretch}{1.3}
\caption{Study 2: The ensemble average of NRMSE, SSIM, and PSNR evaluated on D-type breasts within the testing dataset}
\label{tab:dtype}
\centering
\begin{tabular}{c|c|c}
\hline
 & \multicolumn{2}{c}{U-Net-RT Models} \\
 & Trained on all types & Trained on Non-D types\\
\hline
NRMSE (SD)  & 0.2859 (0.0445) & 0.3269 (0.0627) \\
\hline
SSIM (SD)   & 0.9268 (0.0148) & 0.9240 (0.0159) \\\hline
PSNR (SD) & 32.51 (2.027) & 31.38 (2.276) \\
\hline
\end{tabular}
\parbox{\linewidth}{
\vspace{0.15cm} Note: SD stands for standard deviation. PSNR is measured in dB.
The values of NRMSE, SSIM, and PSNR were evaluated for both Non-D-types and D-type of breasts using two U-Net-RT models, one trained on all types and the other trained only on Non-D types. 
}
\end{table}
\subsection{Study 3 - Impact of rare tumor occurrence within the training dataset} \label{sec:results:tumor}

\subsubsection{Comparison of U-Net-RT with single-channel approaches}
\begin{figure}[htbp]
    \centering
    \includegraphics[width=0.4\textwidth]{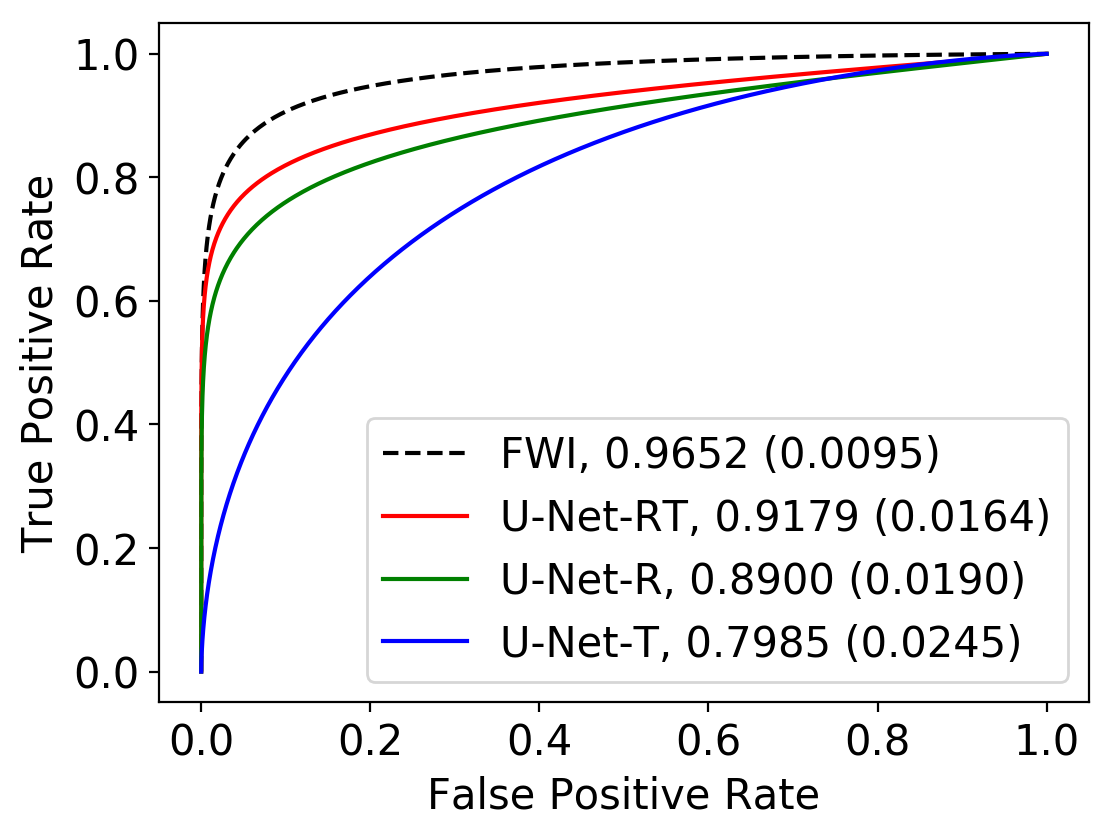}
    \caption{ROC curves and AUC values of the numerical observers using the SOS estimates generated by the U-Net-RT, U-Net-R, U-Net-T, and FWI as inputs. Bracketed numbers indicate the standard error for AUC estimations. }
    \label{fig:roc_case1}
\end{figure}

Consistent with the findings shown in Table \ref{tab:4all}, the U-Net-RT demonstrated superior reconstruction accuracy in both tumor and non-tumor regions compared to the U-Net-R and U-Net-T, as indicated by the NRMSE values in Table \ref{tab:case1}. 
In addition, the ROC curves and AUC values shown in Fig. \ref{fig:roc_case1} demonstrate that the U-Net-RT outperformed both the U-Net-R and U-Net-T in terms of task-based performance.
In particular, as the threshold decreases, the ROC curves of the numerical observers applied to the SOS estimates generated by the U-Net-RT and U-Net-R networks show a slower rise in the false positive rate (FPR, the ratio of false positives to the total number of actual negatives) with a rapid increase in the true positive rate (TPR, the ratio of true positives to the total number of actual positives) compared to the ROC curve corresponding to the U-Net-T reconstruction network.
This highlights that, consistent with the observations in Fig. \ref{fig:unet_results}, the U-Net-T is more prone than the U-Net-R and U-Net-RT to produce hallucinations that can be classified as false positives.
It should be noted that the U-Net-RT shows a marginally higher TPR than the U-Net-R, suggesting improved ability to identify tumors.
This is supported by its greater AUC, which may be attributed to its improved reconstruction accuracy in tumor areas, as measured by NRMSE.
This rapid increase in sensitivity at relatively high levels of specificity for the U-Net-RT highlights a good range of thresholds for a specific clinical application in the case where the potential cost of false positives is high.

\begin{table}[htbp]
\renewcommand{\arraystretch}{1.3}
\caption{Study 3: The ensemble of NRMSE at tumor and non-tumor regions for FWI, U-Net-RT, U-Net-R, U-Net-T, and the family of U-Net-RTw evaluated on the testing dataset}
\label{tab:case1}
\centering
\begin{tabular}{c|c|c}
\hline
 & \multicolumn{2}{c}{NRMSE (SD)} \\
Reconstruction method & Non-tumor regions & Tumor regions \\
\hline
FWI & 0.3315 (0.0844) & 0.3477 (0.0482) \\
U-Net-RT & 0.2195 (0.0685) & 0.4157 (0.2024) \\
U-Net-R & 0.3447 (0.1388) & 0.5542 (0.2468) \\
U-Net-T & 0.5397 (0.1506) & 0.7893 (0.0919) \\
\hline
U-Net-RTw2 & 0.2196 (0.0689) & 0.4119 (0.2122) \\
U-Net-RTw5 & 0.2222 (0.0693) & 0.3802 (0.1591) \\
U-Net-RTw10 & 0.2282 (0.0689) & 0.3719 (0.1595) \\
U-Net-RTw20 & 0.2374 (0.0715) & 0.3589 (0.1382) \\
\hline
\end{tabular}
\parbox{\linewidth}{\vspace{0.15cm} 
    Note: The \textit{p}-values, determined using the Mann-Whitney U test, between NRMSE at tumor regions and NRMSE at non-tumor regions, were below 0.01 except for FWI (p=0.06)
}
\end{table}

One important observation is that the U-Net models exhibited lower reconstruction accuracy in tumor areas compared to non-tumor areas as measured by NMRSE in Table \ref{tab:case1}.
The statistical analysis revealed significant differences between the two areas (\textit{p}<0.01).
In contrast, FWI did not reveal any statistically significant difference in the NRMSE between the two regions, as shown by the non-significant \textit{p}-value (>0.06).
It should be noted that the U-Net-RT showed even higher NRMSE in tumor regions than FWI. 
\begin{figure}[htbp]
    \centering
    \includegraphics[width=0.49\textwidth]{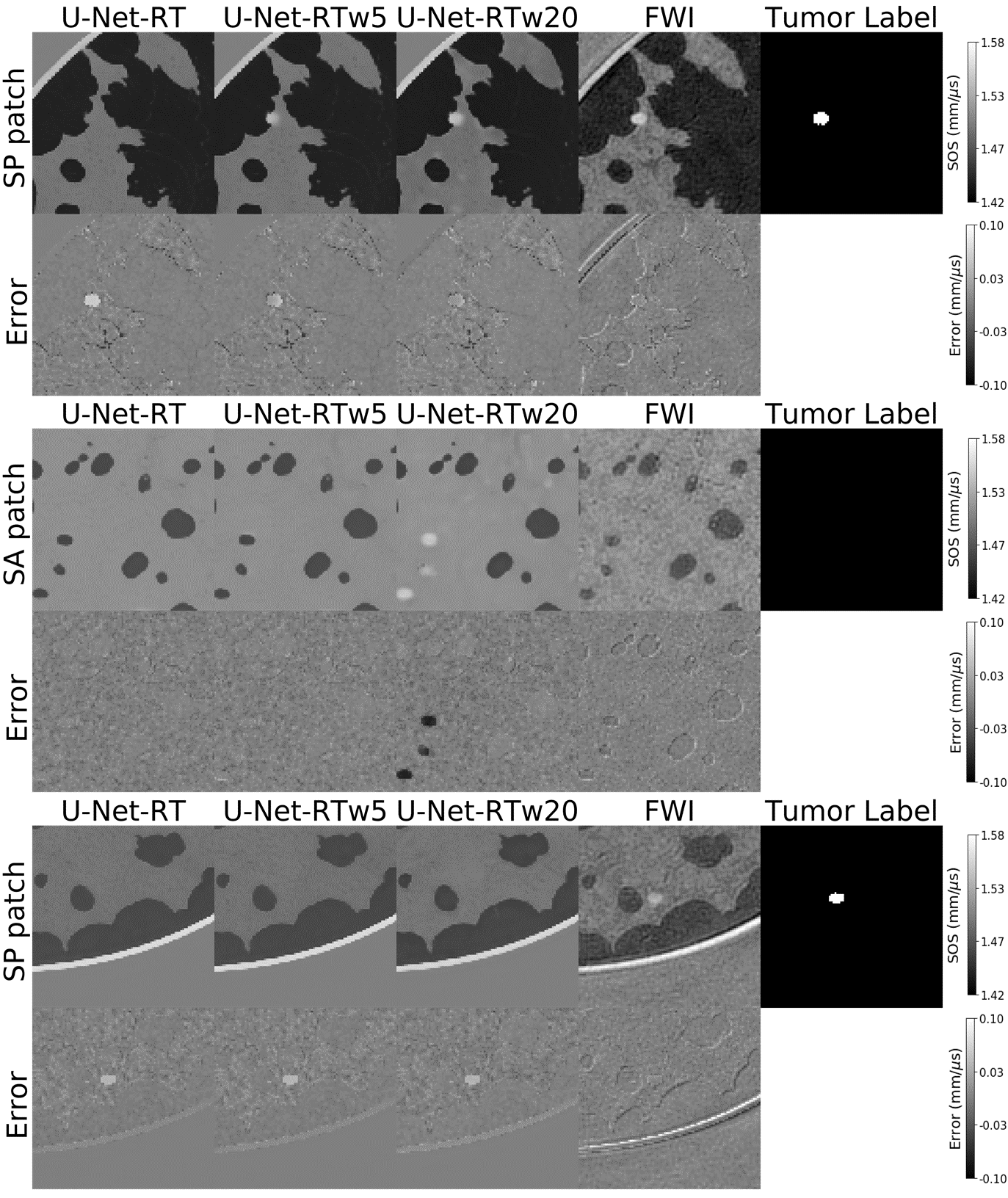}
    \caption{Three examples of 24mm-by-24mm patches from the testing dataset. From left to right: patches obtained from the SOS maps produced by U-Net-RT, U-Net-RTw5, U-Net-RTw20, and FWI, with the final column presenting the corresponding label map indicating tumor presence/absence. For each example, from top to bottom: SP or SA patch followed by the corresponding error map. The first row shows a case where the fine-tuned U-Net-RT models can resolve tumors while the standard U-Net-RT fails. The second row shows a case where the fine-tuned U-Net-RT with high weight values (U-Net-RTw20) can produce false positives. Finally, the third highlights the limited ability of the standard and fine-tuned U-Net-RT models to resolve the tumor when compared to FWI.}
    \label{fig:patch}
\end{figure}

\begin{figure}[htbp]
    \centering
    \includegraphics[width=0.4\textwidth]{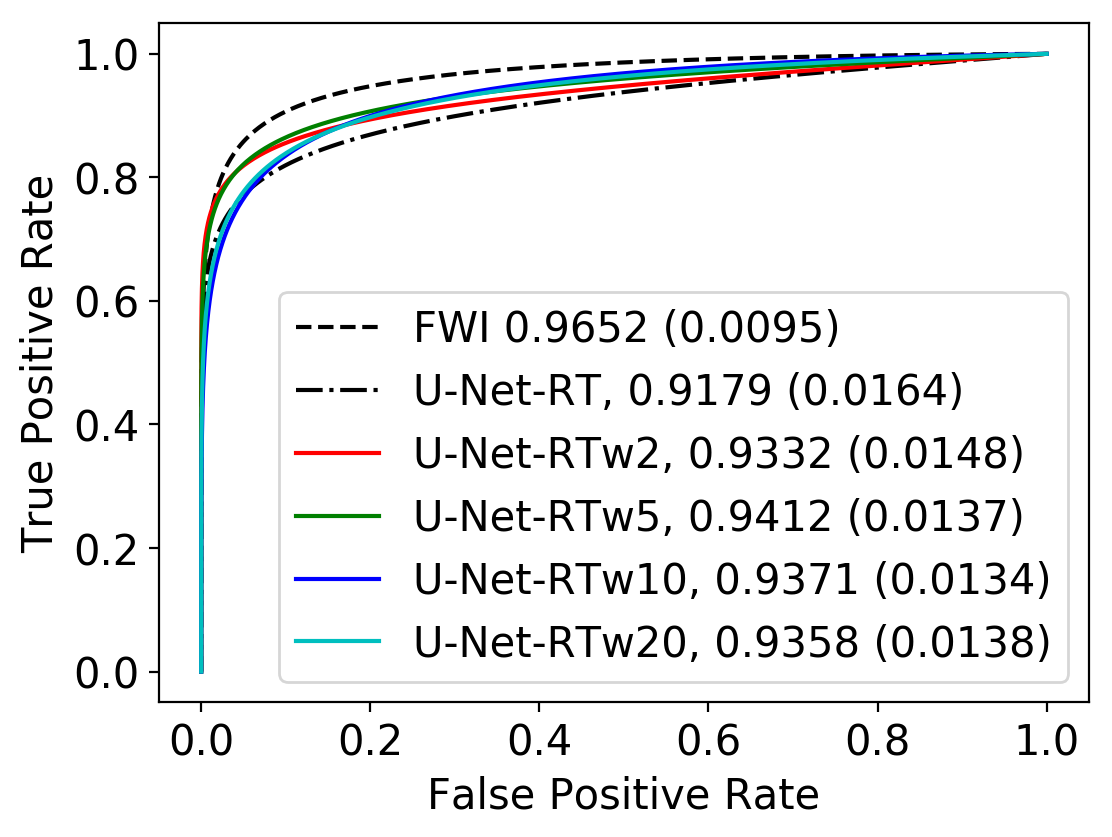}
    \caption{ROC curves for fine-tuned U-Net-RT using WMSE loss function: U-Net-RTw2, U-Net-RTw5, U-Net-RTw10, and U-Net-RTw20, along with FWI and U-Net-RT for comparison. Bracketed numbers indicate the standard error for AUC estimations.}
    \label{fig:roc_case2}
\end{figure}

This reduced reconstruction accuracy in tumor regions of the images produced by the U-Net-RT model resulted in a diminished task-based performance as measured by the CNN-based observer compared to that of FWI.
Specifically, the U-Net-RT yielded a consistently lower TPR across the entire FPR range [0,1] in comparison to the FWI method.
As can be seen in the first and third rows in Fig. \ref{fig:patch}, there are instances where the U-Net-RT cannot resolve tumors, resulting in false negatives by the CNN-based observer, while the FWI method is capable of reconstructing these tumors. 
To address this issue, a fine-tuning approach to increase the reliability of tumor tissues detection, and ultimately improve task-based performance, is assessed next.

\subsubsection{Fine-tuned U-Net-RT using the WMSE loss function}
The results obtained from the fine-tuned U-Net-RT models demonstrated a notable enhancement in reconstruction accuracy inside tumor regions when compared to the standard U-Net-RT model, as measured by NRMSE and reported in Table \ref{tab:case1}.
This improvement is visually demonstrated in the first row of Fig. \ref{fig:patch}: both the U-Net-RTw5 and U-Net-RTw20 revealed the presence of tumors in their SOS estimates, but the standard U-Net-RT did not. 
The improved tumor reconstruction accuracy resulted in slight enhancements in the AUC values for all fine-tuned models, as seen in Fig. \ref{fig:roc_case2}.

Moreover, it was observed that with increasing weight values, the fine-tuned U-Net-RT models achieved a higher level of accuracy in reconstructing tumor areas, as measured by NRMSE. 
This is evident in the first row of Fig. \ref{fig:patch}, where the U-Net-RTw20 estimated the SOS distribution of the tumor more accurately, with decreased errors compared to the U-Net-RTw5.
Conversely, the reconstruction accuracy in non-tumor areas diminishes with increasing the weight value, as shown in Table \ref{tab:case1}. 
This indicates that the use of high-weight values has the potential to introduce false structures in non-tumor areas.
A notable example is found in the estimate produced by the U-Net-RTw20 in the second row of Fig. \ref{fig:patch}.
This highlights a potential trade-off between the accuracies of tumor and non-tumor regions yielded by the fine-tuning approach.

It should be noted that this trade-off is also evident in their ROC curves.
The fine-tuned U-Net-RT models with relatively higher weight values, such as the U-Net-RTw10 and U-Net-RTw20, exhibited decreased sensitivity at high specificity levels as shown in Fig. \ref{fig:roc_case2}, relative to the U-Net-RTw2 and U-Net-RTw5.
These findings are consistent with the data shown in the second row of Fig. \ref{fig:patch}, indicating that the use of greater weights might potentially lead to an increase in the occurrence of false positives. 
On the other hand, the U-Net-RTw10 and U-Net-RTw20 showed increased sensitivity when the FPR was relatively high compared to the U-Net-RTw2 and U-Net-RTw5.
The U-Net-RTw5 performed the best in terms of the AUC value, highlighting the importance of carefully choosing weight values.

Nevertheless, the fine-tuned U-Net-RT models still exhibited lower AUC values compared to FWI.
One potential factor contributing to the suboptimal performance of the fine-tuned U-Net-RT model is its occasional inability to accurately reconstruct certain types of tumors. 
An illustrative example is shown in the third row of Fig. \ref{fig:patch}, where FWI resolved a tumor while the fine-tuned U-Net-RT models yielded false negatives.

\begin{figure*}[!htb]
    \centering
    \includegraphics[width=0.9\textwidth]{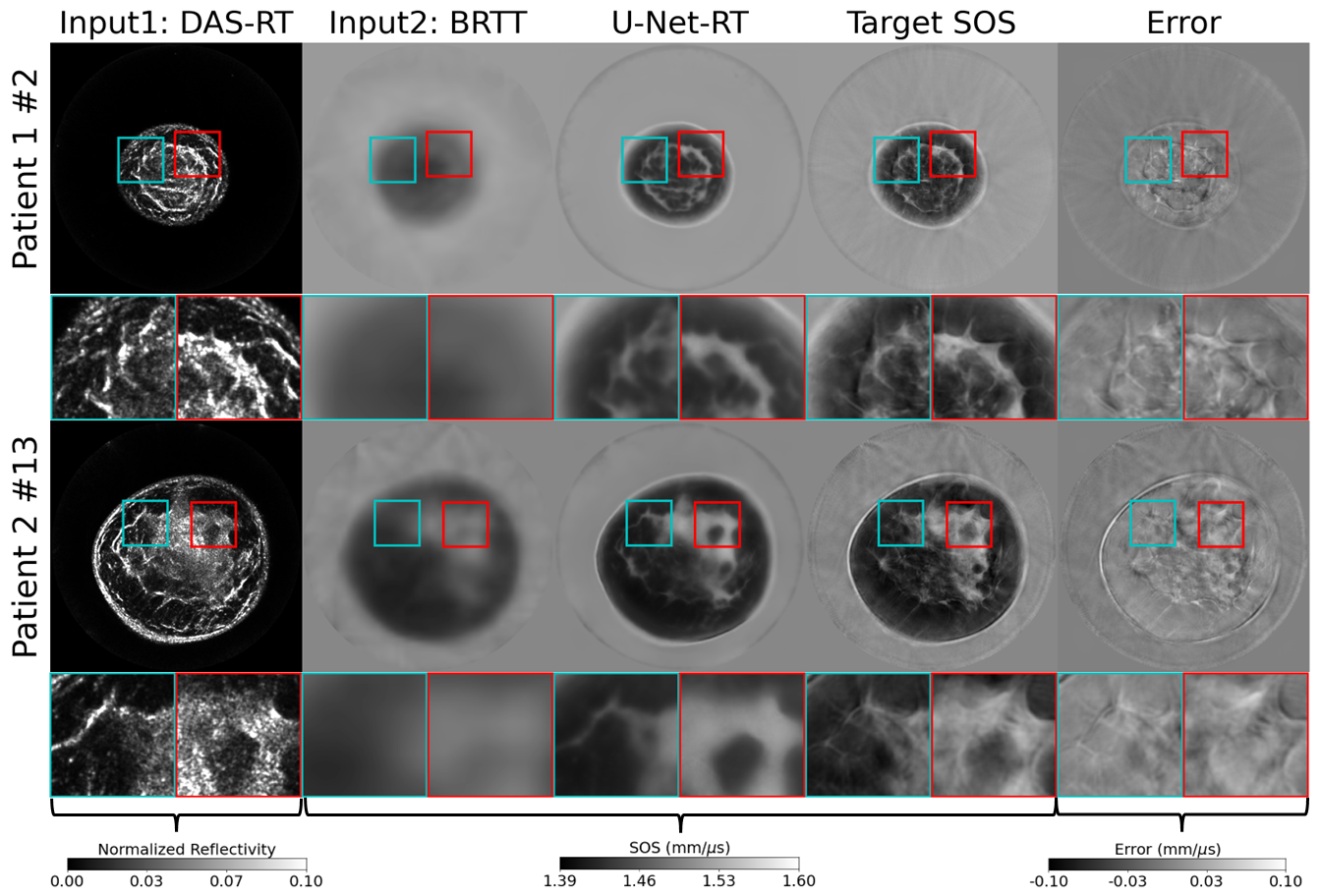}
    \caption{Application of the dual-modality IILR method to  clinical USCT breast data. From left to right: reflectivity amplitude map reconstructed by DAS-RT (scaled using its maximum value), SOS maps reconstructed by BRTT, SOS map produced by U-Net-RT, target SOS map produced by FWI, and the corresponding error map representing the difference between the target SOS map and the SOS map produced by U-Net-RT. Each row represents a slice from a different patient. The rows are labeled as 'Patient X $\#$Y', where X is the patient number and Y is the slice number. Lower Y values indicate slices closer to the nipple. Each image represents a 212 mm-by-212 mm area. Square insets highlight specific patches (37.5mm-by-37.5mm) within the images, with zoomed views shown below each image. DAS-RT and BRTT images, serving as inputs for U-Net-RT, contribute tissue boundary information and background SOS information, respectively, to the SOS map produced by U-Net-RT.}
    \label{fig:unet_results_real}
\end{figure*}

\begin{figure*}[!htb]
    \centering
    \includegraphics[width=0.8\textwidth]{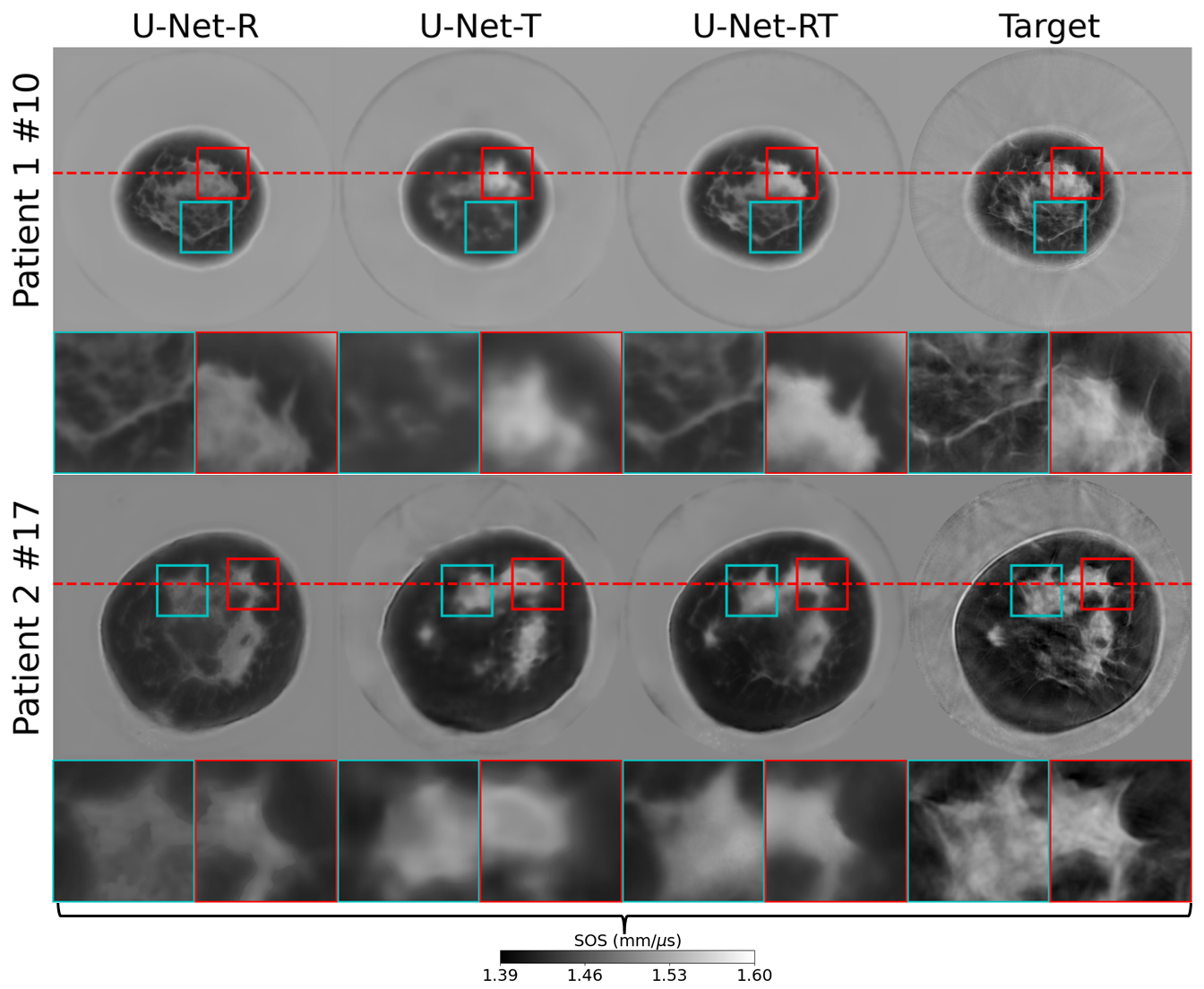}
    \caption{Comparison of the dual-modality and single-modality IILR methods using clinical USCT breast data. From left to right: SOS estimates produced by the U-Net-R, U-Net-T, U-Net-RT, and FWI (target SOS). Each row represents a slice from a different patient. The rows are labeled as 'Patient X $\#$Y', where X is the patient number and Y is the slice number. Lower Y values indicate slices closer to the nipple. Each image represents a 212 mm-by-212 mm area. Square insets highlight specific patches (37.5mm-by-37.5mm) within the images, with zoomed views shown below each image. The red dashed lines in the two testing cases indicate the locations of the line profiles shown in Fig. \ref{fig:line_profile_real}. While U-Net-RT and U-Net-R exhibit similar tissue structures due to the DAS-RT input informing boundary locations and shapes, U-Net-R shows significant bias in high-SOS regions, and U-Net-T is more prone to missing fine details.}
    \label{fig:unet_results_real2}
\end{figure*}

\begin{figure*}[!htb]
    \centering
    \begin{tabular}{cc}
    \includegraphics[width=0.445\textwidth]{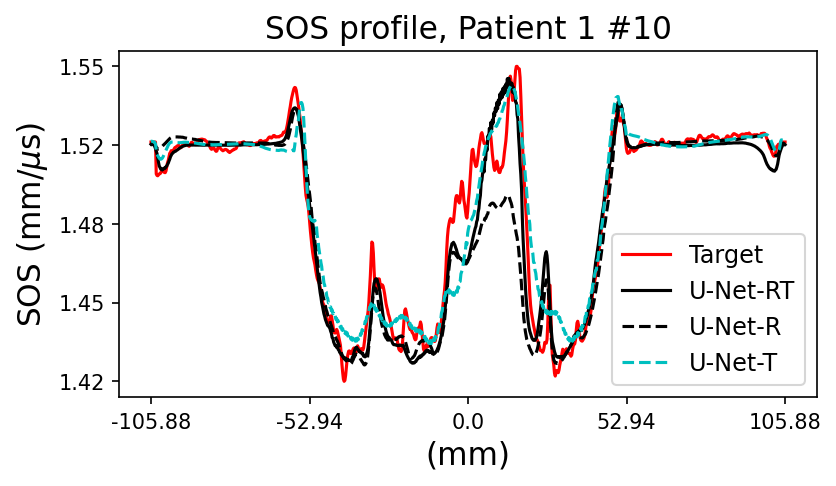} &
    \includegraphics[width=0.445\textwidth]{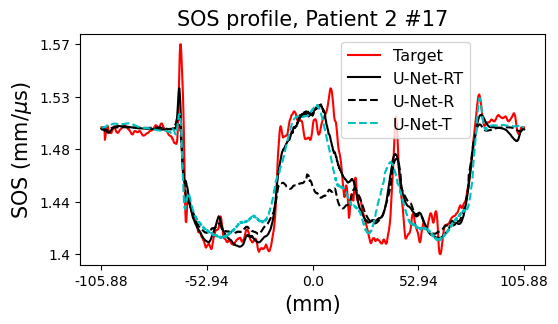} \\
    \multicolumn{1}{c}{(a)} & \multicolumn{1}{c}{(b)}
    \end{tabular}
    \caption{
    Comparison of SOS line profiles for (a) Patient 1 $\#10$ and (b) Patient 2 $\#17$, corresponding to red dashed lines in Fig. \ref{fig:unet_results_real2}. The results include the target SOS produced by FWI and the SOS estimates produced by the U-Net-RT, U-Net-R, and U-Net-T. The U-Net-R exhibits bias in certain high-SOS regions, while the U-Net-T output shows distortion at tissue boundaries when compared to the U-Net-RT results. }
    \label{fig:line_profile_real}
\end{figure*}

\section{Application to \emph{in-Vivo} Experimental Data} \label{invivo}

The previous section presented systematic virtual imaging studies evaluating the impact of BRTT and DAS-RT input images on the quality of SOS estimates produced by the dual-modality IILR method. 
In this section, a preliminary investigation of the dual-modality IILR method is performed using clinical USCT breast data.
\subsection{Experimental setup}
Archived USCT data from previous studies \cite{duric2009detection} were utilized.
The raw waveform data were completely anonymized. 
The USCT system consisted of a ring transducer with a 110 mm radius, comprising 1024 elements arranged around the breast inside a water bath.
Each of the 1024 emitting transducer elements sequentially insonified the breast using an excitation pulse with a center frequency of 2.5 MHz, and all 1024 elements simultaneously measured the received wavefield. 
This sequential acquisition was repeated until each of the 1024 elements served as a source to acquire a full tomographic dataset.
Waveform data were sampled by each transducer element at a 12 MHz rate.
To acquire volumetric data, these 1024 acquisitions were vertically repeated at 2.5 mm intervals, with the ring array translating vertically in a step-wise manner.

\subsection{Training details}
As in the virtual imaging studies, the U-Net-RT model was trained using BRTT and DAS-RT images as inputs and FWI images as target SOS maps. 
Additionally, two single-input models were trained: U-Net-R using only DAS-RT images, and U-Net-T using only BRTT images.
BRTT images were reconstructed using TOF data obtained from raw measurements. 
DAS-RT images were reconstructed using data filtered through a 3 MHz bandpass filter (1.5-4 MHz bandwidth) to ensure clinically relevant features were resolved. 
FWI images were reconstructed using a frequency-domain approach \cite{sandhu2015frequency}. 
The inversion process utilized a frequency range from 370 kHz to 800 kHz, employing a multi-scale strategy that progressed from lower to higher frequencies.
The corresponding BRTT images served as the initial models.
The computational grid for BRTT and FWI image reconstruction consisted of 912-by-912 pixels with a pixel dimension of 0.25 mm. 
DAS-RT images were initially reconstructed at a higher resolution of 1824-by-1824 pixels with a pixel size of 0.125 mm, then downsampled to 912-by-912 pixels using linear interpolation for consistency. 
All images were then cropped from 912-by-912 to 848-by-848 pixels to remove the background water SOS region, consistent with the approach used in the virtual imaging studies described in Section \ref{ns:training data}.

The training, validation, and testing datasets were formed using data from 16 subjects selected from the anonymized archive.
For each subject, slices from left, right, or both breasts were included, depending on the quality of the acquisition and the presence of artifacts in FWI images.
The number of usable slices per subject varied, typically ranging from 10 to 30, due to differences in breast sizes and the presence of implants or other factors affecting image quality.
The 16 subjects were divided into separate sets: 13 for training, 1 for validation, and 2 for testing. 
This division resulted in 337 slices for training, 26 for validation, and 36 for testing, with no subject's data overlapping between sets.
The architecture of the U-Net models and their optimization settings remained consistent with those used in the virtual imaging studies. 
The training process employed the standard MSE as the loss function.

\begin{table}[!htb]
\renewcommand{\arraystretch}{1.3}
\caption{Study 1: The ensemble average of NRMSE, SSIM, and PSNR for U-Net-RT, U-Net-R, U-Net-T, and FWI evaluated on the clinical testing dataset}
\label{tab:real}
\centering
\resizebox{\columnwidth}{!}{%
\begin{tabular}{c|c|c|c}
\hline
 & \multicolumn{3}{c}{Reconstruction Method} \\
\cline{2-4}
 & U-Net-RT & U-Net-R & U-Net-T \\
\hline
NRMSE (SD) &0.2355 (0.0213) & 0.2950 (0.0348)& 0.2748 (0.0235)\\\hline
SSIM (SD) &0.8845 (0.0394) &  0.8742 (0.0401)& 0.8655 (0.0430) \\\hline
PSNR (SD) &28.33 (2.151) & 26.43 (2.499) & 27.01 (2.105) \\\hline
\end{tabular}}
\parbox{\linewidth}{\vspace{0.15cm} 
Note: SD stands for standard deviation. PSNR is measured in dB.}
\end{table}

\subsection{Experimental results}
Examples of the SOS estimates produced by the U-Net-RT on the testing dataset (not present in the training dataset), along with the corresponding input images and error maps, are shown in Fig. \ref{fig:unet_results_real}.
Although not shown here, the quality of the SOS estimates produced by the U-Net-RT on the testing set was comparable that of the training set, indicating that the model generalized well.
As shown in Fig. \ref{fig:unet_results_real}, the U-Net-RT outputs align well with the overall variations captured in the target SOS maps.  However, it can be observed that the U-Net-RT outputs are smoother than the target SOS maps. This reduced sharpness may be partially attributed to differences in structural representations between the target and the DAS-RT input images. 
Some fine-scale features present in FWI but not clearly represented in the DAS-RT images were not fully recovered by U-Net-RT, resulting in smoother reconstructions in these areas.
Interestingly, this smoothing characteristic of U-Net-RT appears to mitigate certain artifacts. 
Specifically, the U-Net-RT was able to filter out some of the cycle-skipping artifacts present in the target SOS maps. 
For instance, artifacts crossing the breast, particularly prominent as abruptly low SOS regions in the cyan patches of the target SOS maps for both patients, were removed in the U-Net-RT outputs. 
The absence of these artifacts in the U-Net-RT outputs can be attributed to their lack of presence in the corresponding BRTT and DAS-RT input maps. 
Furthermore, streak artifacts, predominantly visible in the water bath, were also largely filtered out because the DAS-RT images do not exhibit such artifacts.

To gain further insights into the effect of each input modality, examples of the SOS estimates produced by the U-Net-RT, U-Net-R, and U-Net-T are shown in Fig. \ref{fig:unet_results_real2}, along with the corresponding target SOS. 
Line profiles extracted from these examples are presented in Fig. \ref{fig:line_profile_real}.
Consistent with the virtual imaging studies, tissue structures in the U-Net-RT and U-Net-R outputs were similar.
This similarity indicates that the DAS-RT input informs the location and shape of tissue boundaries in both models. 
In contrast, the U-Net-T output was more prone to missing fine details.
However, the U-Net-R exhibited significant bias in some high-SOS regions.
It should be noted that, as highlighted in the patches of Fig. \ref{fig:unet_results_real}, the U-Net-RT and U-Net-R outputs possessed additional structures---informed by the DAS-RT input---that are not apparent in the corresponding target SOS. 
This difference can be attributed to low-pass filtering in FWI, which emphasizes large-scale structures, while high-pass filtering in the DAS-RT emphasizes thin boundaries and small features.

The ensemble averages of NRMSE, SSIM, and PSNR values evaluated on 36 slices (extracted from the two subjects in the testing set) are presented in Table \ref{tab:real}. 
These metrics are comparable to those obtained in the simulation study, demonstrating U-Net-RT's superiority over U-Net-T and U-Net-R, as well as its general effectiveness when applied to real data.

\section{Discussion}\label{sec:discussion}
While FWI is the standard for accurate high-resolution SOS reconstruction, it often proves computationally intensive for large-scale problems, requiring high-performance computing hardware \cite{lucka2021high}.  This can hinder its application in low-resource settings.
In contrast, the dual-modality IILR method offers a computationally efficient approach providing high-resolution SOS estimates.
The total reconstruction time is comparable to that of either BRTT or DAS-RT image reconstruction, as once trained, the network can rapidly produce an SOS estimate in near real-time.
In scenarios where a further refinement of the SOS estimate is desired, the output of the reconstruction network can serve to initialize a physics-based FWI method. 
Because of the relatively high-quality of the initial SOS distribution, the number of iterations needed to reach a specified level of convergence may be significantly reduced as compared to the standard process in which a BRTT image is employed to initialize the FWI method.

The preliminary study using clinical USCT breast data demonstrated the effectiveness of using the dual-modality inputs, showing potential for practical application. 
However, one limitation of this clinical study relates to the limited size and diversity of the training set, which included only 14 subjects. 
Specifically, while distant slices from the same patient can exhibit significant structural differences, adjacent slices often share similar features, potentially limiting the overall structural diversity in the training set.
This contrasts with the virtual imaging studies, where each slice was extracted from a unique NBP, ensuring greater diversity in the training data. Future work should focus on expanding the training dataset with a larger and more diverse set of clinical cases, which could enhance the model's generalizability and robustness for a wider range of breast anatomies.

The proposed framework offers flexibility in customizing the IILR network to suit certain diagnostic tasks \cite{adler2022task, lozenski2024learned,li2022task}.
By modifying the training loss to include task-related information, the reconstruction network can produce images whose diagnostic utility can be enhanced.
However, despite the use of a fine-tuning strategy with the WMSE loss, a decrease in tumor detection performance compared to FWI was observed. 
This demonstrates that learning-based image reconstruction methods, even when incorporating multiple inputs, may suffer from inherent limitations due to missing information from BRTT and DAS-RT, especially when data-fidelity constraints are not incorporated.

To address this challenge, future research could aim to promote consistency between the network's output and the measurement data during  training, an approach widely adopted in various IILR-based methods for medical imaging \cite{schlemper2017deep, wu2023deep}. 
For the dual-channel IILR method, one potential approach could incorporate a data consistency step using a full wave propagation operator. This operator, implemented via numerical methods such as the pseudospectral k-space method, would be applied to the IILR network output (i.e., SOS distribution) as a post-processing step. The wave propagation operator would enforce physical constraints and data consistency. Unlike end-to-end MILR approaches that attempt to learn the entire physics from measured data alone, this hybrid method can alleviate such burden by leveraging accurate numerical methods for wave propagation. This approach potentially allows for reduced network complexity and the use of standard architectures (such as the U-Net architecture used in this work). However, the effectiveness of such an approach for USCT would require extensive further research and evaluation.

\section{Conclusions}
\label{sec:conclusions}
This work investigated the effectiveness of using TT and RT images as concurrent inputs for an IILR method that seeks to produce high-resolution SOS estimates in USCT. 
Virtual imaging studies demonstrated that this dual-modality method enabled the production of accurate, high-resolution SOS maps.
It also demonstrated robust generalizability when tested on a breast type  not included in the training data.
While FWI outperformed the IILR method with regards to tumor detection, a fine-tuning strategy that assigned higher weights to tumor regions in the loss function narrowed this performance gap.
Preliminary \emph{in-vivo} results indicated the dual-modality IILR method's ability to capture quantitative SOS information and potentially reduce certain artifacts when compared to FWI. 
This study provides insights into the potential and challenges of deep learning methods for USCT reconstruction, advancing high-resolution SOS imaging techniques.

\section*{Acknowledgement}
\color{black}{This research used the Delta advanced computing and data resource which is supported by the National Science Foundation (award OAC 2005572) and the State of Illinois. Delta is a joint effort of the University of Illinois Urbana-Champaign and its National Center for Supercomputing Applications.}
\appendix 
\subsection{Proximal GN method for BRTT with a box constraint}\label{app}
In the GN method, the nonlinear least-squares problem is approximated by a series of linear least-squares subproblems.
At the $j$-iteration, the linearized sub-problem subject to a box constraint is defined as:
\begin{equation}
\label{eq:backgrounds:lsiterative}
\mathbf{y}^{j*} = \argmin_{\mathbf{y}^{j}\in \mathbb{R}^K}\sum_{m=0}^{M-1} \| \mathbf{d}_{m}^j - \mathbf{L}^{\mathbf{b}^j}_m \mathbf{y}^{j} \|_2^2 +I_{B}(\mathbf{b}^j + \mathbf{y}^j)
\end{equation}
where $\mathbf{d}_{m}^j$ is the residual vector given by $\mathbf{t}^{obs}_m - \mathbf{L}_m^{\mathbf{b}^j} \mathbf{b}^{j}$ and $I_{B}$ is an indicator function for a convex set $B\subset \mathbb{R}^N$ which is given by:
\begin{align}
    I_{B}(\mathbf{b}) = 
    \begin{cases}
    0 &\text{\qquad if\,\,} \mathbf{b} \in B \\
    \infty &\text{\qquad if\,\,} \mathbf{b} \notin B\\
    \end{cases}
\end{align}
where $B$ is specifically defined as follow for $\forall \mathbf{b} \in B$:
\begin{equation}
\begin{cases}
    \mathbf{b}[i] =0 &\text{\qquad if \,\,} \|\mathbf{r}_i -
\mathbf{r}_c \|_2 > r_{FOV}\\
    b_{min}\le \mathbf{b}[i] \le b_{max} &\text{\qquad if \,\,} \|\mathbf{r}_i -
\mathbf{r}_c \|_2 \le r_{FOV}\\
\end{cases}
\end{equation}
where $\mathbf{r}_c$ is the grid point at the center, $r_{FOV}$ is the radius of imaging FOV, and $b_{min}$ and $b_{max}$ are the desired lower- and upper bounds of the elements of $\mathbf{b}$. 
In this study, $b_{\text{min}}$ and $b_{\text{max}}$ are set as $\frac{1}{1.6}\mu\text{s}/\text{mm}$ and $\frac{1}{1.4}\mu\text{s}/\text{mm}$.

The solution to the proximal sub-problem at $j$-th iteration $\Delta \mathbf{b}^{j*}$ corresponds to the step that the GN method takes to update the current estimate. 
Specifically, the estimate at iteration $j+1$ is given by \begin{equation}
     \mathbf{b}^{j+1} = \mathbf{b}^{j} + \mu^j\mathbf{y}^{j*},
\end{equation}
where the quantity $\mu^j$ is the step-size at the $j$-th iteration.
The proximal gradient method can be used to solve the sub-problem \eqref{eq:backgrounds:lsiterative}. The details of the algorithm are described in \ref{alg:pgn-box}.

\begin{algorithm}[htb]
  \caption{Proximal GN Method with a Box Constraint}
  \label{alg:pgn-box}
  \begin{algorithmic}[1]
    \REQUIRE TOF data $\{\mathbf{t}^{obs}_m\}_{m=0}^{N_e-1}$, initial slowness $\mathbf{b}^0$ and initial slowness perturbation $\Delta\mathbf{b}^0$
    \FOR{$j = 0,1,\dots$}
      \STATE Solve the Eikonal equation to obtain the prediction of TOF $\{\mathbf{T}_m^{j}\}_{m=0}^{N_e-1}$
      \STATE Construct the ray-tracing matrices $\{\mathbf{L}^{\mathbf{b}^j}_m\}_{m=0}^{N_e-1}$
      \STATE Evaluate  $\{\mathbf{d}^{j}_{m}\}_{m=0}^{N_e-1}$
      \STATE Evaluate the gradient of the data fidelity term
      \begin{align*}
        \mathbf{g}^j &= -\sum_{m=0}^{N_e-1}{{\mathbf{L}^{\mathbf{b}^j}_m}^T \mathbf{d}^{j}_{m}}
      \end{align*}
      \STATE Evaluate the proximal gradient
      \begin{align*}
        \mathbf{g}^j_{\text{prox}} &= \mathbf{b}^j - \text{prox}_{I_{B}} (\mathbf{b}^j - \mathbf{g}^j)
      \end{align*}
      \IF{$\|\mathbf{g}^j_{\text{prox}}\|<\text{TOL}$}
        \STATE \textbf{return} $\mathbf{b}^j$
      \ENDIF
      \STATE Solve the sub-problem
      \begin{align*}
        \mathbf{y}^j &= \argmin_{\mathbf{y}\in\mathbb{R}^K} \frac{1}{2}\sum_{m=0}^{N_e-1}\|\mathbf{L}^{\mathbf{b}^j}_m\mathbf{y}-\mathbf{d}^j\|^2_2 + I_B({\mathbf{b}^j+\mathbf{y}})
      \end{align*}
      \STATE Define $\lambda^j = {\mathbf{g}^j}^T \mathbf{y}^j$
      \STATE Find $\mu^j\le 1$ such that $\mathbf{b}^{j+1} = \mathbf{b}^j + \mu^j\mathbf{y}^j$ satisfies the sufficient descent condition:
      \begin{align*}
        \sum_{m=0}^{N_e-1}\|\mathbf{t}^{obs}_m - \mathbf{L}^{\mathbf{b}^{j+1}}_m \mathbf{b}^{j+1}\|^2_2 \le &\sum_{m=0}^{N_e-1}\|\mathbf{t}^{obs}_m - \mathbf{L}^{\mathbf{b}^j}_m \mathbf{b}^j\|^2_2 \\
        &+ \mu^j\lambda^j
      \end{align*}
    \ENDFOR
  \end{algorithmic}
\end{algorithm}
Here, the proximal operator of the function $I_{B}$, $\text{prox}_{I_B}$, is given by
\begin{equation}
    \text{prox}_{I_B}(\mathbf{b}) = \argmin_{\mathbf{x}\in B} \left(\frac{1}{2}\| \mathbf{x}-\mathbf{b}\|^2_2\right) 
\end{equation}

\bibliographystyle{IEEEtran}
\bibliography{usct}

\end{document}